# Site Preference and Possible Coexistence of Antiferromagnetic Order and Magnetic Frustration in $(Co_{1-x}Mg_x)_{10}Ge_3O_{16}$ (0 ≤ x ≤ 30%)

*Gina Angelo,[a] Qiang Zhang,[b] Dylan Correll[a] and Xin Gui[a]\**

[a] Department of Chemistry, University of Pittsburgh, Pittsburgh, PA, 15260, USA
[b] Neutron Scattering Division, Oak Ridge National Laboratory, Oak Ridge, TN, 37831, USA
Address correspondence to E-mail: xig75@pitt.edu

## *Abstract*

Geometrically frustrated magnetism has attracted tremendous attention while chemical doping has been utilized as an important tool to probe frustrated magnetism in various systems. Here we perform a systematic study by doping non-magnetic $Mg^{2+}$ into a magnetically complicated system, $Co_{10}Ge_3O_{16}$, which contains three frustrated sublattices of $Co^{2+}$, e.g., triangular Co1, Kagome Co2 and Co3 sublattices. By growing crystals for $(Co_{1-x}Mg_x)_{10}Ge_3O_{16}$ (0 < x ≤ 30%), we observed obvious site preference of $Mg^{2+}$ on Co1 and Co3 sites over the Co2 site. Powder X-ray diffraction (XRD) patterns confirm the high purity of the samples and indicate systematic peak shift, consistent with the loading compositions. Although previously investigated, the magnetic structure and expected magnetic frustration in this system are not fully uncovered. Our temperature-dependent magnetic susceptibility measurements suggest that the high-temperature magnetostructural phase transition with antiferromagnetic ordering and a low-temperature broad peak are suppressed with $Mg^{2+}$ doping, while two new magnetic features emerge at high $Mg^{2+}$ level. Moreover, the structural phase transition from high-temperature *R*-3*m* to low-temperature *C*2/*m* space group is absent at the antiferromagnetic ordering temperature, as confirmed by single-crystal XRD. By analyzing the heat capacity and neutron powder diffraction results of the highest doped sample, $(Co_{0.7}Mg_{0.3})_{10}Ge_3O_{16}$, we speculate that the Co1 site is responsible for the long-range antiferromagnetic ordering, while the other two sites are short-range correlated in addition to a $Mg^{2+}$-induced spin-glass state. This study provides more insights into the complex magnetism in $Co_{10}Ge_3O_{16}$ by using the non-magnetic $Mg^{2+}$ as a probe. However, detailed magnetic structure requires further efforts on growing large single crystals.

## *Introduction*

Magnetically frustrated materials have been extensively studied due to the exotic quantum phenomena they can potentially host, including but not limited to quantum spin liquid/ice,[1–8] Majorana fermion,[1,9–15] topological order,[3,5,16] and unconventional superconductivity[17–22]. One of the most common origins of magnetic frustration is geometric frustration in triangular,[8,23,24] Kagome,[25–28] pyrochlore[29–33] and square[34,35] lattices of magnetic species, while an unambiguous observation of a quantum spin liquid/ice remains an unsolved problem.[5,23,36] From a material design perspective, this can be partially solved by design and discovery of new geometrically frustrated compounds with magnetic atoms/ions. In this regard, chemical disorder is usually utilized as a tool to tune magnetic exchange interactions in a promising structural motif with geometric frustration and lead to magnetic frustration, e.g., the disorder-induced/suppressed spin liquid states in many compounds were reported.[37–40]

Spinels, formulated as $AB_2X_4$ with a B-site pyrochlore sublattice in a *Fd-3m* space group, are one of the most widely investigated systems for frustrated magnetism, e.g., $ZnFe_2O_4$ with potential quantum-spin-liquid-like behavior based on its unconventional exchange pathways and frustrated $GeCo_2O_4$ with unusually frustration mechanisms.[41–43] Relevant to spinels, $Co_{10}Ge_3O_{16}$ was reported as an intergrowth between alternating $GeCo_2O_4$ spinel slabs and rock-salt-type layers.[44] It crystallizes in the space group of *R-3m* under room temperature while a structural phase transition to *C2/m* occurs at ~203 K accompanied by a long-range antiferromagnetic ordering.[45,46] Three geometrically frustrated $Co^{2+}$ (S = 3/2) sublattices exist in $Co_{10}Ge_3O_{16}$ including one triangular and two Kagome lattices. Complex low-temperature magnetism was proposed in $Co_{10}Ge_3O_{16}$ that it undergoes another ferrimagnetic transition and potential spin reorientation below ~20 K.[45] However, detailed magnetic structure was not reported, likely due to the structural complexity and the resolution of neutron powder diffraction.[45]

What motivates this research is the intriguing geometrically frustrated lattices of $Co^{2+}$ in $Co_{10}Ge_3O_{16}$. One would expect strong magnetic frustration existing in $Co_{10}Ge_3O_{16}$, which was, however, not uncovered in previous studies. In this paper, we employ the non-magnetic cation, $Mg^{2+}$, to probe the magnetism in $Co_{10}Ge_3O_{16}$ by synthesizing and characterizing $(Co_{1-x}Mg_x)_{10}Ge_3O_{16}$ (0 ⩽ x ⩽ 30%). The chemical compositions are confirmed by both single-crystal and powder X-ray diffraction. $Mg^{2+}$ shows an obvious site preference. Studies on magnetic properties suggest a suppression of the magnetostructural phase transition with $Mg^{2+}$ doping. At x = 30%, the structural phase transition is no longer observed at the antiferromagnetic transition temperature. Moreover, additional magnetic features at high $Mg^{2+}$ concentration are observed, which are believed to be short-range correlation/ordering and a doping-induced spin-glass state. Based on heat capacity and neutron powder diffraction measurements on $(Co_{0.7}Mg_{0.3})Ge_3O_{16}$, we provide possible explanations for the complex magnetism in $Co_{10}Ge_3O_{16}$ and Mg-doped ones. Moreover, signs of magnetic frustrations are observed coexisting with the long-range antiferromagnetic ordering. Our study sets $Co_{10}Ge_3O_{16}$ as a promising platform for investigating the interplay between long-range magnetic ordering, magnetic frustration and geometric frustration.

*Experimental Details*

**Synthesis of Polycrystalline $(Co_{1-x}Mg_x)_{10}Ge_3O_{16}$:** Polycrystalline $(Co_{1-x}Mg_x)_{10}Ge_3O_{16}$ (x = 0%, 1%, 5%, 10%, 15%, 20%, and 30%) was prepared by mixing $Co_3O_4$ (99.7%, ~400 mesh, Thermo Scientific) powder, $GeO_2$ (99.999%, powder, Thermo Scientific) powder, and MgO (99.99%, powder, Thermo Scientific) in stoichiometric ratios with a 2% excess of $Co_3O_4$. Excess $Co_3O_4$ was necessary to achieve high purity in all samples. The starting materials were placed in an alumina crucible and heated in air to 1150 °C for 48 h several times with intermittent grindings until high purity was achieved.

**Growth of Single-Crystalline $(Co_{1-x}Mg_x)_{10}Ge_3O_{16}$:** Crystals of $(Co_{1-x}Mg_x)_{10}Ge_3O_{16}$ (x = 1%, 5%, 10%, 15%, 20%, and 30%) were grown using $K_2MoO_4$ flux. Powder $(Co_{1-x}Mg_x)_{10}Ge_3O_{16}$ (synthesized as described above) was added to an alumina crucible with $K_2MoO_4$ (98%, powder, Sigma Aldrich) in a 3.5% wt ratio. The samples were heated in air to 1200 °C for 12 h then slowly cooled at 1 °C/h to 1050 °C before cooling to room temperature at 180 °C/h. Excess flux was washed away with DI water and the resulting crystals were allowed to dry overnight before characterization.

**Single Crystal and Powder X-Ray Diffraction (XRD):** The crystal structure of $(Co_{1-x}Mg_x)_{10}Ge_3O_{16}$ was determined using a Bruker D8 Quest ECO diffractometer equipped with APEX5 software and Mo radiation ($\lambda_{K\alpha}$=0.71073 Å). The crystals were mounted on a Kapton loop with STP oil. $(Co_{0.7}Mg_{0.3})_{10}Ge_3O_{16}$ at 107 K was determined using a Bruker D8 Venture Duo diffractometer equipped with APEX5 software and Mo radiation ($\lambda_{K\alpha}$=0.71073 Å). Both the direct method and fill-matrix least-squares on $F^2$ procedure within the SHELXTL package were used to solve the crystal structures.[47,48] The phase purity of $(Co_{1-x}Mg_x)_{10}Ge_3O_{16}$ was determined using a Bruker D2 PHASER with Cu radiation ($\lambda_{K\alpha}$=1.54060 Å, Ge monochromator). The Bragg angle was measured from 5 to 100° at a rate of 1.7°min$^{-1}$ with a step of 0.012°.

**Physical Property Measurement:** The Quantum Design Dynacool Physical Property (PPMS) was used to measure the DC magnetization on polycrystalline samples from 2 K to 300 K under various applied external magnetic fields using the equipped ACMS II option. Field-dependent magnetization data were collected in the range of -9 to 9 T under different temperatures. Temperature dependence of magnetic susceptibility was measured from 2 to 300 K under external magnetic field. Zero field cool (ZFC) and field cool (FC) were both performed. Heat capacity was measured using a standard relaxation method in the PPMS. Magnetic properties were performed on as-synthesized powder sample.

Heat capacity measurement was conducted using pelletized samples after being annealed at the reaction temperature for 10 hours.

**Neutron Powder Diffraction:** Neutron powder diffraction was conducted at high-resolution time-of-Flight powder diffractometer POWGEN, located at SNS, ORNL. A powder sample of $(Co_{1-x}Mg_x)_{10}Ge_3O_{16}$ (x=30%) with mass of 1.25 g was loaded in a vanadium can with 6 mm diameter. Helium exchange gas was filled in to ensure good thermal conductivity. A POWGEN Automatic Changer (PAC) was adopted as sample environment to cover the temperature region from 8 to 300 K. Two neutron frames with center wavelengths of 1.5 and 2.665 Å were used to collect data, covering different regions of $d$-spacing. Only the results obtained from wavelength of 1.5 Å is shown in this paper. Measurements were performed at 300, 100 and 8 K.

## Results and Discussion

**Crystal Structure and Mg Distribution in $(Co_{1-x}Mg_x)_{10}Ge_3O_{16}$:** The crystal structures of $(Co_{1-x}Mg_x)_{10}Ge_3O_{16}$ (x = 1%, 5%, 10%, 15%, 20% and 30%) determined by single crystal X-ray diffraction (XRD) at room temperature appear to be consistent with the reported structure,[44–46] all of which crystallize in a rhombohedral space group *R*-3*m* (S.G. 166). The crystallographic data including atomic sites, site occupancies, refined anisotropic displacement parameters and equivalent isotropic thermal displacement parameters of all compounds are summarized in Table 1 and Tables S1 and S2 in the supporting information (SI). The refined doping value x are 0.7(3)%, 5(1)%, 9(1)%, 14(1)%, 20(2)% and 29(2)%, which are consistent with the loading composition for synthesizing polycrystalline samples. Therefore, we will use the loading compositions in the context below for simplicity and clarity.

The crystal structure of $(Co_{1-x}Mg_x)_{10}Ge_3O_{16}$ is shown in Figure 1a, which consists of three $Co^{2+}$ sites (Co1, Co2 and Co3) and two $Ge^{4+}$ sites (Ge1 and Ge2). The Ge1 and Ge2 sites are six- and four-coordinated with $O^{2-}$ while all $Co^{2+}$ sites are six-coordinated and forming Co@$O_6$ octahedron. Additionally, Co2@$O_6$ and Ge1@$O_6$ are located within the same *ab* plane, whereas Co1@$O_6$ and Ge2@$O_6$ reside in a different *ab* plane. Co3@$O_6$ octahedra, however, form a distinct *ab* plane of their own. Three Co sublattices can be found in $(Co_{1-x}Mg_x)_{10}Ge_3O_{16}$, including Co1 (triangular; Wyckoff 3*b*; interlayer separation of ~9.6 Å), Co2 (Kagome; Wyckoff 9*e*; interlayer separation of ~9.6 Å) and Co3 (Kagome; Wyckoff 18*h*; interlayer separation of ~4.7 Å & ~4.9 Å), as shown in Figure 1b. The $Co^{2+}$-$Co^{2+}$ bond lengths are ~5.95 Å, ~2.98 Å and ~2.92 Å for Co1, Co2 and Co3, respectively.

It is noteworthy that all three Co@$O_6$ octahedra are slightly distorted from ideal geometry in different ways. Table 2 summarizes all the unique $Co^{2+}$-$O^{2-}$ bond lengths and bond angles based on the nomenclature provided in Figure 1c such that $O_{ap.}$ and $O_{eq.}$ ($d_{ap.}$ and $d_{eq.}$) represent $O^{2-}$ ($Co^{2+}$-$O^{2-}$ bond) located along the apical and equatorial directions while bond angle $\delta$ stands for the $O_{eq.}$-Co-$O_{ap.}$ angle towards the positive *c* axis. It can be seen that Co1@$O_6$ is the least distorted octahedron with identical $d_{ap.}$ and $d_{eq.}$ and $\delta$ slightly bigger than 90° (90.9° – 91.1°). Moreover, Co2@$O_6$ octahedron exhibits the smallest distortion angle (89.2° – 89.7°) with $Mg^{2+}$ doping and a slightly elongated octahedron, while Co3@$O_6$ shows the largest distortion angle (91.7° – 92.0°) and compressed octahedron. Overall, $Mg^{2+}$ doping does not significantly alter the ($Co^{2+}$,$Mg^{2+}$)@$O_6$ octahedra.

Based on the single crystal XRD results, the occupancy of $Mg^{2+}$ on all three $Co^{2+}$ sites are summarized in Figure 1d. When x = 1%, $Mg^{2+}$ is only observed on Co2 site with an occupancy of 2.2(1.1)%. Interestingly, when x increases to 5%, an obvious site preference is seen on Co3 (6(1)%) and Co1 (4.9(1.9)%) sites over Co2 site (2.9(1.3)%), which persists up to x = 30% where the occupancy of $Mg^{2+}$ on Co1/Co2/Co3 is 28.2(1)%/17.2(7)%/35.2(5)%.

**Analysis of Phase Purity:** Powder XRD was employed to determine the phase purity of $(Co_{1-x}Mg_x)_{10}Ge_3O_{16}$. As shown in Figure 2a, the powder XRD patterns of $(Co_{1-x}Mg_x)_{10}Ge_3O_{16}$ are consistent with what was reported for undoped $Co_{10}Ge_3O_{16}$[45] and show high purity for all samples, except for the undoped $Co_{10}Ge_3O_{16}$ which contains minor impurity of $Co_3O_4$. Therefore, physical properties were only measured for Mg-doped samples. The inset of Figure 2a shows the systematic shift of the powder XRD patterns by varying x. The (110) and (0012) peaks represent the changes on the *a/b* axis and *c* axis. The observed shift towards larger 2θ indicates the shrinking of *a* and *c* due to the smaller ionic radius of $Mg^{2+}$ (0.72 Å) compared to $Co^{2+}$ (0.745 Å).[49] The results of Le Bail fitting further prove the shrinkage of the unit cell with $Mg^{2+}$ doping, as can be seen in Figure 2b. From x = 0% to 30%, the length of *a* and *c*, and the volume of the unit cell decrease monotonically.

**Magnetic Properties:** Magnetic properties measurements of $(Co_{1-x}Mg_x)_{10}Ge_3O_{16}$ were conducted on as-synthesized polycrystalline samples for x ≥ 1%. Figure 3 illustrates the temperature-dependent magnetic susceptibility (χ) and $χ^{-1}$ for both undoped and Mg-doped $Co_{10}Ge_3O_{16}$ from 2 K to 380 K under external magnetic field of 0.1 T using zero-field-cooling (ZFC) protocol, while the curves using field-cooling (FC) protocol are shown in Figure S1 in the SI. Consistent with what was reported for $Co_{10}Ge_3O_{16}$,[45] a sharp peak is observed in all Mg-doped compounds at $T_1$ ($T_1$ = 203 K for undoped $Co_{10}Ge_3O_{16}$). $T_1$ decreases monotonically when x increases, i.e., from $T_N$ ~ 202 K for x = 1% to $T_N$ ~ 140 K for x = 30%. $T_1$ was determined to be an antiferromagnetic order along with a structural phase transition from high-temperature *R-3m* to low-temperature *C2/m* symmetry.[45] Moreover, a broad feature corresponding to $T_b$ is seen just below $T_1$ for x ≤ 20%, which then disappears when x = 30%, as illustrated in Figure 3. Such a broad peak is characteristic of materials with (quasi-)two-dimensional (2D) magnetism.[50] However, $T_b$ is either suppressed or submerged into the low-temperature increase of χ when x = 30% below 100 K. Considering the three $Co^{2+}$ sublattices in $Co_{10}Ge_3O_{16}$, it is reasonable to believe that 2D magnetic correlations exist in all $(Co_{1-x}Mg_x)_{10}Ge_3O_{16}$ and thus $T_b$ is buried under the upward χ vs T curve in $(Co_{0.7}Mg_{0.3})_{10}Ge_3O_{16}$, as discussed later. At even lower temperatures, a sudden increase of χ can be found for all samples, which was believed to be a spin reorientation based on ref. 45. Additionally, a cusp is seen for x ≤ 5% at $T_2$ ~10 K (x = 0%),[45] ~6 K (x = 1%) and ~2.8 K (x = 5%), respectively, likely correspond to a second ferrimagnetic-like transition.[45] Interestingly, a new magnetic feature that was not reported emerges at T* for x = 20% (~8.5 K) and 30% (~14.9 K), which will be discussed shortly, while the ferrimagnetic-like transition at $T_2$ is fully suppressed within the measurement range for the two samples.

Based on the linearity of the $χ^{-1}$ vs T curves, Curie-Weiss (CW) fitting is applied to all samples under both ZFC and FC protocols from 300 K to 380 K by using the formula

$$\chi = \frac{C}{T - \theta_{CW}}$$

where C is a temperature-independent constant and is related to the effective moment ($\mu_{eff}$) via $\mu_{eff}/Co = \sqrt{8\frac{C}{n}}$, and $\theta_{CW}$ is the CW temperature. $\mu_{eff}$ is calculated with respect to Co content where n is the number of Co atoms in a formula unit of a sample. The fitted $\theta_{CW}$ ranges from +50 K (x = 1%) to -27 K (x = 30%), indicating a ferromagnetic-to-antiferromagnetic crossover on the magnetic interaction between x = 10% and x = 15%. Note that the $\theta_{CW}$ was reported to be +39.3 K for $Co_{10}Ge_3O_{16}$.[45] Two possible reasons can be attributed to the discrepancy: 1. The inclusion of $Mg^{2+}$ increases the diamagnetic response of the sample, thus introducing a negative $χ_0$ to the system thereby lowering χ and increasing fitted $\theta_{CW}$; 2. The reported $Co_{10}Ge_3O_{16}$ contains 1 mol% of $GeCo_2O_4$ impurity,[45] which may affect the measured χ. Meanwhile, the magnitude of $\mu_{eff}$/Co is consistent with high-spin $Co^{2+}$ when considering the contribution from orbital angular momentum and comparable to the reported value for $Co_{10}Ge_3O_{16}$ (~4.26 $\mu_B$/Co).[45] Additionally, it is noteworthy that the fitted $\theta_{CW}$ or their absolute values are significantly lower than the observed magnetic transition temperatures ($T_1$) for all samples. We attribute this to the fact that octahedral $Co^{2+}$ possesses tremendous magnetic anisotropy such that CW law, assuming isotropic mean-field average of exchange interactions, does not accurately describe the real magnetic interaction between $Co^{2+}$ ions.

As mentioned above, a new magnetic feature is seen under T*~14.9 K when x = 30%. To further demonstrate the magnetic properties of $(Co_{0.7}Mg_{0.3})_{10}Ge_3O_{16}$, Figure 4a presents its temperature-dependent χ and $χ^{-1}$ under an external magnetic field of 0.1 T from 2 K to 380 K with both ZFC and FC protocols. A slight deviation is observed between ZFC and FC curves below ~ 120 K. Such a deviation becomes significant below 40 K where FC curve shows an upturn trend, but ZFC curve exhibits the opposite and forms a broad peak at T*. Because the deviation between ZFC and FC curves is typical for conventional spin glass, the temperature-dependent AC magnetic susceptibility of $(Co_{0.7}Mg_{0.3})_{10}Ge_3O_{16}$ was measured between 5 K and 30 K under the applied magnetic field of 10 Oe for both DC and AC and various frequencies including 193 Hz, 268 Hz, 373 Hz, 518 Hz, 720 Hz, 1001 Hz, 1389 Hz, 1930 Hz, 2684 Hz, 3725 Hz, 5182 Hz, 7200 Hz and 9984 Hz. As shown in Figure 4b, a broad peak was observed at $T^*_1$ ~ 16 K, which is consistent with the T* ~ 14.9 K observed in Figure 3. The minor difference is due to the varied DC magnetic field. $T^*_1$ does not show any frequency-dependent behavior. However, a kink is obviously seen at the highest AC frequency at $T^*_2$ ~ 25 K while it declines when the frequency decreases. For clarity, the first derivatives of $\chi_{AC}'$ ($d(\chi_{AC}')/dT$) were plotted in Figure S2 in the SI to demonstrate the monotonic reduction of $T^*_2$ with lowered frequencies. Such frequency-dependent behavior is a sign for spin-glass state and was observed in other spin-glass systems.[51] Given the geometrically frustrated $Co^{2+}$ sublattices and the extensive substitutional disorders of $Mg^{2+}$ on these sublattices, a $Mg^{2+}$-induced spin-glass state is a reasonable speculation for $T^*_2$. For $T^*_1$, we attribute it to short-range magnetic order between $Co^{2+}$ due to the possible magnetic frustration and the interruption of the magnetic exchange pathways caused by $Mg^{2+}$ doping, as demonstrated later.

**Heat Capacity of $(Co_{0.7}Mg_{0.3})_{10}Ge_3O_{16}$:** To further investigate the magnetic properties of $(Co_{0.7}Mg_{0.3})_{10}Ge_3O_{16}$, temperature-dependent heat capacity ($C_p$) measurement was conducted under no external magnetic field from 2 K to 200 K on a sample pellet sintered at 1150 °C for 12 hours. Figure 5a shows the $C_p$ (T) curves which exhibit a broad feature below ~19 K and a λ-shape transition at ~140 K. The high-temperature transition is consistent with the antiferromagnetic transition reported in ref. 45 for $Co_{10}Ge_3O_{16}$ and what was observed in the magnetic properties for $(Co_{0.7}Mg_{0.3})_{10}Ge_3O_{16}$ at $T_1$. The low-temperature transition possesses a distinct feature compared to undoped $Co_{10}Ge_3O_{16}$. For the undoped $Co_{10}Ge_3O_{16}$, a sharp upturn can be seen below ~15 K with minor entropy change in $C_p/T$ *vs* T curve, which is attributed to spin reorientation or a minor structural change.[45] However, the $C_p/T$ *vs* T curve for $(Co_{0.7}Mg_{0.3})_{10}Ge_3O_{16}$ in the right inset of Figure 5b shows a broad peak starting at ~20 K and centered at ~10 K, likely corresponding to $T^*_1$.

The total heat capacity of a magnetic material in the paramagnetic region can be interpreted as the sum of electronic, phononic and magnonic contributions, as for $C_p = C_{el} + C_{ph} + C_{mag}$. The $C_{el}$ term is determined to be zero due to the insulating behavior of these materials. Here we employ the Debye model for phononic contribution and perform fitting using the formula

$$C_p = 9nR[(\frac{T}{\theta_D})^3 \int_0^{\theta_{D1}/T} \frac{x^4 e^x}{(e^x - 1)^2} dx]$$

where n is the number of atoms per formula unit, R is gas constant, $\theta_D$ is the Debye temperature. The fitting is shown in Figure 5a, which is in good agreement with the high temperature data (160 K – 200 K), while discrepancies exist below ~150 K, indicating the release of magnetic entropy ($\Delta S_{mag}$). Note that the volume per formula unit of the high-temperature (*R-3m*) and low-temperature (*C2/m*) structures are nearly identical (295.197 Å$^3$ *vs* 295.326 Å for $Co_{10}Ge_3O_{16}$). Therefore, the high-temperature $C_p$ fitted from Debye model is approximately identical to the low-temperature one. After subtracting $C_{ph}$ from $C_p$, the residual heat capacity can be attributed to magnetic origins. As shown in Figure 5b, three peaks spanning from 2 K to ~160 K are seen for $C_{mag}/T$. The one at ~140 K is clearly corresponding to the transition at $T_1$ while the peak at ~10 K originates from $T^*_1$ and/or $T^*_2$. An unexpected broad peak shows up at ~54 K, which is believed to be related to $T_b$ that is buried under the upward $\chi(T)$ curve in $(Co_{0.7}Mg_{0.3})_{10}Ge_3O_{16}$. The $C_{mag}/T$ curve is integrated and $\Delta S_{mag}/Co$ is obtained. The $\Delta S_{mag}/Co$ is saturated at ~ 4.2 J/mol/K, which is only ~37% of the expected value for S = 3/2 $Co^{2+}$ (~11.5 J/mol/K), indicating that not all of the $Co^{2+}$ sites are long-range ordered. Given the sharp feature of the transition at $T_1$ and the broadness of $T_b$ and $T^*_1/T^*_2$, it is plausible that only the transition at $T_1$ corresponds to a long-range phase transition, which is consistent with the magnetostructural transition in ref. 45, while $T_b$ and $T^*$ both correspond to short-range ordering.

**Neutron Powder Diffraction of $(Co_{0.7}Mg_{0.3})_{10}Ge_3O_{16}$:** To better interpret the magnetic behavior of $(Co_{0.7}Mg_{0.3})_{10}Ge_3O_{16}$, neutron powder diffraction (NPD) was performed on POWGEN at Oak Ridge National Laboratory. Figure 6a shows the NPD patterns under 8 K, 100 K and 300 K with Rietveld refinement. Consistent with XRD results, no additional nuclear peaks can be observed in NPD patterns, indicating high sample purity and identical crystal structure. It is noteworthy that the low-temperature crystal structure with *C2/m* space group was employed to fit the NPD patterns at 8 K and 100 K. However, no good fitting can be obtained and the refinement does not converge. Thus, we performed single-crystal XRD measurement for $(Co_{0.7}Mg_{0.3})_{10}Ge_3O_{16}$ under 107 K, i.e., below its $T_1$, and determined that $(Co_{0.7}Mg_{0.3})_{10}Ge_3O_{16}$ crystallizes in *R-3m* space group instead of *C2/m*, as shown in Table 3. Therefore, all low-temperature NPD patterns are fitted using *R-3m* space group.

At low Q region, at least six new peaks at Q ~1.0, ~1.23, ~1.44, ~2.33, ~2.57 and ~2.79 Å$^{-1}$ corresponding to a commensurate ***k*** = (00 3/2) and an increase in intensity for Q ~ 1.64 Å$^{-1}$ can be seen, suggesting long-range magnetic ordering, which is consistent with the observed $T_1$ and ref. 45. Moreover, an additional peak at Q ~ 0.6 Å$^{-1}$ was reported for undoped $Co_{10}Ge_3O_{16}$,[45] which is likely originated from the transition at $T_2$. However, this peak is absent in $(Co_{0.7}Mg_{0.3})_{10}Ge_3O_{16}$, consistent with the observed suppression of $T_2$ in Figure 3. Additionally, the peak intensity of all the peaks marked with asterisks, i.e., magnetic peaks, in Figure 6b increase when cooling from 100 K to 8 K. This is strong evidence for the spin-canted antiferromagnetism generated from the same antiferromagnetic ordering that occurs at $T_1$. Detailed magnetic structure is not solved in this work based on NPD due to the structural complexity and will be reported in another paper based on single crystal neutron diffraction. But possible magnetic space groups of this system are shown in Figure S3.[52]

**Analysis of High-Temperature Magnetic Ordering and Low-Temperature Magnetic Frustration in $(Co_{0.7}Mg_{0.3})_{10}Ge_3O_{16}$ and Beyond:** Based on the results above, here we can propose the possible magnetic ordering type of $(Co_{0.7}Mg_{0.3})_{10}Ge_3O_{16}$ and provide guidance for future magnetic structure solutions of this system via neutron single crystal diffraction if larger crystals can be grown. According to Figure 5b, the entropy change of $T^*_1/T^*_2$, $T_b$ and $T_1$ are estimated to be ~0.5 J/mol/K, ~2.7 J/mol/K and 1.0 J/mol/K, respectively. The expected $\Delta S_{mag}/Co$ for a fully ordered high-spin octahedral $Co^{2+}$ is ~11.5 J/mol/K. Therefore, the observed $\Delta S_{mag}/Co$ are ~4.3%, ~23.5% and ~8.7% of the expected value, meaning that there are ~4.3%, ~23.5% and ~8.7% of the total number of $Co^{2+}$ that participated in releasing the magnetic entropy. On the other hand, based on Figure 1d, we can obtain that the number of $Co^{2+}$ on Co1, Co2 and Co3 sites are ~10%, ~35% and ~55% of the total amount of $Co^{2+}$ per unit cell, respectively. Given that $T_1$ must be a long-range antiferromagnetic ordering temperature, i.e., the whole corresponding Co site needs to be magnetically ordered, it can be speculated that the Co1 site ordered antiferromagnetically below $T_1$, while Co2 and Co3 sites contribute to the broad peaks at $T_b$ and $T^*$ in Figure 5b. Because of their short-range nature, it is hard to attribute Co2/Co3 to either peak.

However, based on their 2-D lattice geometries, we can conclude that in addition to the observed spin-glass state at $T^*_2$, another geometry-induced magnetic frustration exists at low-temperatures (< ~100 K) for Co2 and Co3 sites when Co1 possesses long-range antiferromagnetic ordering below $T_1$. Given the absence of sharp transitions in the reported heat capacity of undoped $Co_{10}Ge_3O_{16}$ below 100 K,[45] it can be speculated that same short-range order and magnetic frustration exist in $Co_{10}Ge_3O_{16}$, as well as the Mg-doped $Co_{10}Ge_3O_{16}$ investigated in this paper. Moreover, as shown in Figure 9 in ref. 45, the newly emerged (001) peak at 5 K in neutron powder diffraction of $Co_{10}Ge_3O_{16}$ possesses a larger full width at half maximum compared to the nuclear peak at (00 1/2). Thus, this new peak at 5 K that is likely corresponding to $T_2$ may be induced by short-range antiferromagnetic ordering instead of a new magnetic phase transition, which is also evidenced by the lack of transition in heat capacity of $Co_{10}Ge_3O_{16}$ shown in Figure 8 in ref. 45. For the ferrimagnetic-like transition in Figure 3, i.e. the increase of $\chi$ at low temperatures, it is likely corresponding to both the static moments on Co2/Co3 sites and the potentially canted low-temperature antiferromagnetism on Co1 site. This can also explain the coercivity existing in $(Co_{1-x}Mg_x)_{10}Ge_3O_{16}$ as detailed in next section. Similar behavior can be found in magnetically frustrated $RbSb_3Mn_9O_{19}$.[53] A diagram sketching the proposed evolution of magnetic order/interaction in $Co_{10}Ge_3O_{16}$ and Mg-doped ones is shown in Figure S4. It is necessary to point out that a short-range ordered system is expected to show diffuse scattering peaks in NPD patterns, which is absent in $(Co_{0.7}Mg_{0.3})_{10}Ge_3O_{16}$. We attribute the lack of feature to the fact that non-magnetic $Mg^{2+}$ disrupts magnetic exchange interactions between $Co^{2+}$ so that the correlation length is too short to be detected by our NPD measurements. Therefore, $Co_{10}Ge_3O_{16}$ can potentially be a system with alternatingly stacked antiferromagnetic (Co1) and magnetically frustrated (Co2 and Co3) layers. However, growth of larger crystals is necessary to resolve detailed magnetic structure.

**Field-Dependence of Magnetization:** The hysteresis loops of $(Co_{1-x}Mg_x)_{10}Ge_3O_{16}$ under 2 K are illustrated in Figure 7a. With increasing dopant concentration, hysteresis loops exhibit systematic changes. The magnetization is normalized to the number of $Co^{2+}$ per formula unit and its value at 9 T ($M_{9T}$) decreases with increasing x, i.e., from 1.0 $\mu_B/Co^{2+}$ when x = 1% to 0.9/0.8/0.75/0.68/0.54 $\mu_B/Co^{2+}$ when x = 5%/10%/15%/20%/30%. The lowered $M_{9T}$ with increasing $Mg^{2+}$ concentration is consistent with what is discussed above. At low temperatures, the system undergoes a crossover between long-range ferrimagnetic order, perhaps coexisting with spin reorientation, for small x and short-range antiferromagnetic interactions for large x. Therefore, under a finite external magnetic field, it becomes more difficult for the system to reach saturated moments. The hysteresis loops of $(Co_{1-x}Mg_x)_{10}Ge_3O_{16}$ under other temperatures are shown in Figure S5 in the SI where systematic changes in $M_{9T}$ are also seen.

In addition to the varied $M_{9T}$, it is also observed that the virgin curves from 0 T to 9 T for x ≤ 20% are significantly distinct from the rest of the hysteresis loops. This is usually due to the irreversible movement of magnetic domain wall, however there are other possibilities. Zhao et al. explain a similar anomalous virgin curve via kinetics of a first-order transition which can be slowed by rapid change in environment, in this case, competing AFM and FM phases.[54] Meanwhile, metamagnetic transitions are observed in all samples. For clarity, the first derivatives of the virgin curves are plotted in Figure 7b. Two metamagnetic transitions can be seen for x ≤ 20%, while the transition fields increase for both with more $Mg^{2+}$. For x = 30%, only one metamagnetic transition can be observed between 0 T and 9 T, which is likely to be the same lower-field transition seen in other samples. The metamagnetic transitions become broader with increasing x, which can be attributed to the increasing concentration of non-magnetic impurity ($Mg^{2+}$) existing in the magnetic exchange pathways.

It was found that the coercivity of $(Co_{1-x}Mg_x)_{10}Ge_3O_{16}$ can be changed based on batch size. Figure S6 shows the hysteresis loops of $(Co_{1-x}Mg_x)_{10}Ge_3O_{16}$ from a different batch with smaller scale, i.e., 200 mg in total compared to 2 grams in total for which we have performed all the other measurements. Both batches were synthesized using identical methods. It can be observed that the coercivity of 200-mg batch is much smaller compared to the 2-gram batch in these samples. Interestingly, the virgin curves are nearly identical, excluding the fact that the differences originate from impurities. It can be speculated that the coercivity of $(Co_{1-x}Mg_x)_{10}Ge_3O_{16}$ is highly sensitive to the crystallite size and regular high-temperature solid-state synthesis with different batch sizes can tune the coercivity.

## *Conclusion*

In this paper, we utilize non-magnetic $Mg^{2+}$ to probe the complex magnetism of $Co_{10}Ge_3O_{16}$. By growing the single crystals of $(Co_{1-x}Mg_x)_{10}Ge_3O_{16}$ (0 < x ≤ 30%), we determined that $Mg^{2+}$ showed an obvious preference to occupy Co3 and Co1 sites over Co2 site. By measuring the magnetic properties of $(Co_{1-x}Mg_x)_{10}Ge_3O_{16}$, we observed a suppression of the magnetostructural transition from ~203 K to

~140 K, as well as the suppression of the two-dimensional-like broad peak in magnetic susceptibility. Moreover, the low-temperature transition observed in $Co_{10}Ge_3O_{16}$ becomes absent with $Mg^{2+}$ doping, while two new transitions occur at high $Mg^{2+}$ concentration (x = 20% and 30%). Additionally, the structural phase transition no longer exists for x = 30% at the antiferromagnetic transition temperature. By analyzing the heat capacity and neutron powder diffraction patterns of the highest-doped sample, $(Co_{0.7}Mg_{0.3})_{10}Ge_3O_{16}$, we speculate that Co1 is the long-term antiferromagnetically ordered site, while the other two sites are short-range correlated/ordered in addition to a doping-induced spin-glass state. However, further investigation into the detailed magnetic structure of this system is necessary when larger crystals can be grown.

## *Acknowledgements*

Research at the University of Pittsburgh is supported by the startup funds for X.G. A portion of this research used resources at the Spallation Neutron Source, as appropriate, a DOE Office of Science User Facility operated by the Oak Ridge National Laboratory. The beam time was allocated to POWGEN on proposal number IPTS-36328.1.

## *Appendix A. Supplementary data*

Supplementary data to this article can be found online at xxxxxxx.

Atomic coordinates, equivalent isotropic displacement parameters, and anisotropic thermal displacement parameters for $(Co_{1-x}Mg_x)_{10}Ge_3O_{16}$. Magnetic susceptibility of $(Co_{1-x}Mg_x)_{10}Ge_3O_{16}$ under field-cooling protocol. First derivatives of AC magnetic susceptibility of $(Co_{0.7}Mg_{0.3})_{10}Ge_3O_{16}$. Possible magnetic space groups from neutron powder diffraction. Proposed evolution of magnetic ordering/correlation in $Co_{10}Ge_3O_{16}$ and $(Co_{1-x}Mg_x)_{10}Ge_3O_{16}$. Hysteresis loops of $(Co_{1-x}Mg_x)_{10}Ge_3O_{16}$ under various temperatures. Comparison of hysteresis loops between the 200-mg and 2-gram batches.

## References

(1) Takagi, H.; Takayama, T.; Jackeli, G.; Khaliullin, G.; Nagler, S. E. Concept and Realization of Kitaev Quantum Spin Liquids. *Nat Rev Phys* **2019**, *1*, 264–280.
(2) Clark, L.; Abdeldaim, A. H. Quantum Spin Liquids from a Materials Perspective. *Annu. Rev. Mater. Res.* **2021**, *51*, 495–519.
(3) Chamorro, J. R.; McQueen, T. M.; Tran, T. T. Chemistry of Quantum Spin Liquids. *Chem. Rev.* **2021**, *121*, 2898–2934.
(4) Gingras, M. J. P.; McClarty, P. A. Quantum Spin Ice: A Search for Gapless Quantum Spin Liquids in Pyrochlore Magnets. *Rep. Prog. Phys.* **2014**, *77*, 056501.

(5) Broholm, C.; Cava, R. J.; Kivelson, S. A.; Nocera, D. G.; Norman, M. R.; Senthil, T. Quantum Spin Liquids. *Science* **2020**, *367*, eaay0668.

(6) Knolle, J.; Moessner, R. A Field Guide to Spin Liquids. *Annu. Rev. Condens. Matter Phys.* **2019**, *10*, 451–472.

(7) Balents, L. Spin Liquids in Frustrated Magnets. *Nature* **2010**, *464*, 199–208.

(8) Anderson, P. W. Resonating Valence Bonds: A New Kind of Insulator? *Mater. Res. Bull.* **1973**, *8*, 153–160.

(9) Kasahara, Y.; Ohnishi, T.; Mizukami, Y.; Tanaka, O.; Ma, S.; Sugii, K.; Kurita, N.; Tanaka, H.; Nasu, J.; Motome, Y.; Shibauchi, T.; Matsuda, Y. Majorana Quantization and Half-Integer Thermal Quantum Hall Effect in a Kitaev Spin Liquid. *Nature* **2018**, *559*, 227–231.

(10) Knolle, J.; Kovrizhin, D.L.; Chalker, J.T.; Moessner, R. Dynamics of a Two-Dimensional Quantum Spin Liquid: Signatures of Emergent Majorana Fermions and Fluxes. *Phys. Rev. Lett.* **2014**, *112*, 207203.

(11) Yoshitake, J.; Nasu, J.; Motome, Y. Fractional Spin Fluctuations as a Precursor of Quantum Spin Liquids: Majorana Dynamical Mean-Field Study for the Kitaev Model. *Phys. Rev. Lett.* **2016**, *117*, 157203.

(12) Do, S.-H.; Park, S.-Y.; Yoshitake, J.; Nasu, J.; Motome, Y.; Kwon, Y. S.; Adroja, D. T.; Voneshen, D. J.; Kim, K.; Jang, T.-H.; Park, J.-H.; Choi, K.-Y.; Ji, S. Majorana Fermions in the Kitaev Quantum Spin System α-RuCl$_3$. *Nat. Phys* **2017**, *13*, 1079–1084.

(13) Gordon, J. S.; Catuneanu, A.; Sørensen, E. S.; Kee, H.-Y. Theory of the Field-Revealed Kitaev Spin Liquid. *Nat Commun.* **2019**, *10*, 2470.

(14) Zhou, Y.; Kanoda, K.; Ng, T.K. Quantum Spin Liquid States. *Rev. Mod. Phys.* **2017**, *89*, 025003.

(15) Kitaev, A. Anyons in an Exactly Solved Model and Beyond. *Ann. Phys.* **2006**, *321*, 2–111.

(16) Wen, X. G. Mean-Field Theory of Spin-Liquid States with Finite Energy Gap and Topological Orders. *Phys. Rev. B* **1991**, *44*, 2664–2672.

(17) Roppongi, M.; Ishihara, K.; Tanaka, Y.; Ogawa, K.; Okada, K.; Liu, S.; Mukasa, K.; Mizukami, Y.; Uwatoko, Y.; Grasset, R.; Konczykowski, M.; Ortiz, B. R.; Wilson, S. D.; Hashimoto, K.; Shibauchi, T. Bulk Evidence of Anisotropic S-Wave Pairing with No Sign Change in the Kagome Superconductor CsV$_3$Sb$_5$. *Nat. Commun.* **2023**, *14*, 667.

(18) Tazai, R.; Yamakawa, Y.; Onari, S.; Kontani, H. Mechanism of Exotic Density-Wave and beyond-Migdal Unconventional Superconductivity in Kagome Metal AV$_3$Sb$_5$ (A = K, Rb, Cs). *Sci. Adv.* **2022**, *8*, eabl4108.

(19) Wu, X.; Schwemmer, T.; Müller, T.; Consiglio, A.; Sangiovanni, G.; Di Sante, D.; Iqbal, Y.; Hanke, W.; Schnyder, A.P.; Denner, M.M.; Fischer, M.H. Nature of unconventional pairing in the kagome superconductors AV$_3$Sb$_5$ (A= K, Rb, Cs). *Phys. Rev. Lett.* **2021**, *127*, 177001.

(20) Kiesel, M. L.; Platt, C.; Thomale, R. Unconventional Fermi Surface Instabilities in the Kagome Hubbard Model. *Phys. Rev. Lett.* **2013**, *110*, 126405.

(21) Wang, W.-S.; Li, Z.Z.; Xiang, Y.Y; Wang, Q.H. Competing Electronic Orders on Kagome Lattices at van Hove Filling. *Phys. Rev. B* **2013**, *87*, 115135.

(22) Kiesel, M. L.; Thomale, R. Sublattice Interference in the Kagome Hubbard Model. *Phys. Rev. B* **2012**, *86*, 121105.

(23) Li, Y.; Liao, H.; Zhang, Z.; Li, S.; Jin, F.; Ling, L.; Zhang, L.; Zou, Y.; Pi, L.; Yang, Z.; Wang, J.; Wu, Z.; Zhang, Q. Gapless Quantum Spin Liquid Ground State in the Two-Dimensional Spin-1/2 Triangular Antiferromagnet YbMgGaO$_4$. *Sci. Rep.* **2015**, *5*, 16419.

(24) Man, H.; Halim, M.; Sawa, H.; Hagiwara, M.; Wakabayashi, Y.; Nakatsuji, S. Spin-Orbital Entangled Liquid State in the Copper Oxide $Ba_3CuSb_2O_9$. *J. Phys.: Condens. Matter* **2018**, *30*, 443002.

(25) Dun, Z.L.; Trinh, J.; Li, K.; Lee, M.; Chen, K.W.; Baumbach, R.; Hu, Y.F.; Wang, Y.X.; Choi, E.S.; Shastry, B.S.; Ramirez, A.P. Magnetic ground states of the rare-earth tripod kagome lattice $Mg_2RE_3Sb_3O_{14}$ (RE= Gd, Dy, Er). *Phys. Rev. Lett.* **2016**, *116*, 157201.

(26) Sanders, M. B.; Krizan, J. W.; Cava, R. J. $RE_3Sb_3Zn_2O_{14}$ (RE = La, Pr, Nd, Sm, Eu, Gd): A New Family of Pyrochlore Derivatives with Rare Earth Ions on a 2D Kagome Lattice. *J. Mater. Chem. C* **2016**, *4*, 541–550.

(27) Sanders, M. B.; Baroudi, K. M.; Krizan, J. W.; Mukadam, O. A.; Cava, R. J. Synthesis, Crystal Structure, and Magnetic Properties of Novel 2D Kagome Materials $RE_3Sb_3Mg_2O_{14}$ (RE = La, Pr, Sm, Eu, Tb, Ho): Comparison to $RE_3Sb_3Zn_2O_{14}$ Family. *Phys. status solidi (b)* **2016**, *253*, 2056–2065.

(28) Mendels, P.; Bert, F. Quantum Kagome Antiferromagnet $ZnCu_3(OH)_6Cl_2$. *J. Phys. Soc. Jpn.* **2010**, *79*, 011001.

(29) Gao, B.; Chen, T.; Tam, D. W.; Huang, C.-L.; Sasmal, K.; Adroja, D. T.; Ye, F.; Cao, H.; Sala, G.; Stone, M. B.; Baines, C.; Verezhak, J. A. T.; Hu, H.; Chung, J.-H.; Xu, X.; Cheong, S.-W.; Nallaiyan, M.; Spagna, S.; Maple, M. B.; Nevidomskyy, A. H.; Morosan, E.; Chen, G.; Dai, P. Experimental Signatures of a Three-Dimensional Quantum Spin Liquid in Effective Spin-1/2 $Ce_2Zr_2O_7$ Pyrochlore. *Nat. Phys.* **2019**, *15*, 1052–1057.

(30) Sibille, R.; Gauthier, N.; Lhotel, E.; Porée, V.; Pomjakushin, V.; Ewings, R. A.; Perring, T. G.; Ollivier, J.; Wildes, A.; Ritter, C.; Hansen, T. C.; Keen, D. A.; Nilsen, G. J.; Keller, L.; Petit, S.; Fennell, T. A Quantum Liquid of Magnetic Octupoles on the Pyrochlore Lattice. *Nat. Phys.* **2020**, *16*, 546–552.

(31) Sibille, R.; Lhotel, E.; Pomjakushin, V.; Baines, C.; Fennell, T.; Kenzelmann, M. Candidate quantum spin liquid in the $Ce^{3+}$ pyrochlore stannate $Ce_2Sn_2O_7$. *Phys. Rev. Lett.* **2015**, *115*, 097202.

(32) Fennell, T.; Deen, P. P.; Wildes, A. R.; Schmalzl, K.; Prabhakaran, D.; Boothroyd, A. T.; Aldus, R. J.; McMorrow, D. F.; Bramwell, S. T. Magnetic Coulomb Phase in the Spin Ice $Ho_2Ti_2O_7$. *Science* **2009**, *326*, 415–417.

(33) Morris, D. J. P.; Tennant, D. A.; Grigera, S. A.; Klemke, B.; Castelnovo, C.; Moessner, R.; Czternasty, C.; Meissner, M.; Rule, K. C.; Hoffmann, J.-U.; Kiefer, K.; Gerischer, S.; Slobinsky, D.; Perry, R. S. Dirac Strings and Magnetic Monopoles in the Spin Ice $Dy_2Ti_2O_7$. *Science* **2009**, *326*, 411–414.

(34) Ramirez, A. P. Strongly Geometrically Frustrated Magnets. *Annu. Rev. Mater. Res.* **1994**, *24*, 453–480.

(35) Tsirlin, A.A.; Nath, R.; Abakumov, A.M.; Furukawa, Y.; Johnston, D.C.; Hemmida, M.; Krug von Nidda, H.A.; Loidl, A.; Geibel, C.; Rosner, H. Phase separation and frustrated square lattice magnetism of $Na_{1.5}VOPO_4F_{0.5}$. *Phys. Rev. B* **2011**, *84*, 014429.

(36) Rao, X.; Hussain, G.; Huang, Q.; Chu, W. J.; Li, N.; Zhao, X.; Dun, Z.; Choi, E. S.; Asaba, T.; Chen, L.; Li, L.; Yue, X. Y.; Wang, N. N.; Cheng, J.-G.; Gao, Y. H.; Shen, Y.; Zhao, J.; Chen, G.; Zhou, H. D.; Sun, X. F. Survival of Itinerant Excitations and Quantum Spin State Transitions in $YbMgGaO_4$ with Chemical Disorder. *Nat Commun* **2021**, *12*, 4949.

(37) Sibille, R.; Lhotel, E.; Ciomaga Hatnean, M.; Nilsen, G. J.; Ehlers, G.; Cervellino, A.; Ressouche, E.; Frontzek, M.; Zaharko, O.; Pomjakushin, V.; Stuhr, U.; Walker, H. C.; Adroja, D. T.; Luetkens, H.; Baines, C.; Amato, A.; Balakrishnan, G.; Fennell, T.; Kenzelmann, M. Coulomb Spin Liquid in Anion-Disordered Pyrochlore $Tb_2Hf_2O_7$. *Nat Commun* **2017**, *8*, 892.

(38) Furukawa, T.; Miyagawa, K.; Itou, T.; Ito, M.; Taniguchi, H.; Saito, M.; Iguchi, S.; Sasaki, T.; Kanoda, K. Quantum spin liquid emerging from antiferromagnetic order by introducing disorder. *Phys. Rev. Lett.* **2015**, *115*, 077001.

(39) Ma, Z.; Dong, Z.Y.; Wu, S.; Zhu, Y.; Bao, S.; Cai, Z.; Wang, W.; Shangguan, Y.; Wang, J.; Ran, K.; Yu, D. Disorder-induced spin-liquid-like behavior in kagome-lattice compounds. *Phys. Rev. B* **2020**, *102*, 224415.

(40) Savary, L.; Balents, L. Disorder-Induced Quantum Spin Liquid in Spin Ice Pyrochlores. *Phys. Rev. Lett.* **2017**, *118*, 087203.

(41) Diaz, S.; De Brion, S.; Chouteau, G.; Canals, B.; Simonet, V.; Strobel, P. Magnetic frustration in the spinel compounds $GeCo_2O_4$ and $GeNi_2O_4$. *Phys. Rev. B* **2006**, 74, 092404.

(42) Tomiyasu, K.; Kamazawa, K. A Spin Molecule Model for Geometrically Frustrated Spinel $ZnFe_2O_4$. *J. Phys. Soc. Jpn.* **2011**, *80*, SB024.

(43) Kamazawa, K.; Tsunoda, Y.; Kadowaki, H.; Kohn, K. Magnetic neutron scattering measurements on a single crystal of frustrated $ZnFe_2O_4$. *Phys. Rev. B* **2003**, *68*, 024412.

(44) Barbier, J. $Co_{10}Ge_3O_{16}$. *Acta Cryst C* **1995**, *51*, 343–345.

(45) Barton, P.T.; Seshadri, R.; Llobet, A.; Suchomel, M.R. Magnetostructural transition, metamagnetism, and magnetic phase coexistence in $Co_{10}Ge_3O_{16}$. *Phys. Rev. B* **2013**, *88*, 024403.

(46) Udod, L. V.; Petrakovskiĭ, G. A.; Vorotynov, A. M.; Bayukov, O. A.; Velikanov, D. A.; Kartashev, A. V.; Bovina, A. F.; Shvedenkov, Yu. G.; Baran, M.; Szymczak, R. Magnetic Properties of Aerugite $Co_{10}Ge_3O_{16}$. *Phys. Solid State* **2007**, *49*, 500–504.

(47) Walker, N.; Stuart, D. An Empirical Method for Correcting Diffractometer Data for Absorption Effects. *Acta Cryst A* **1983**, *39*, 158–166.

(48) Sheldrick, G. M. Crystal Structure Refinement with SHELXL. *Acta Cryst C* **2015**, *71*, 3–8.

(49) Shannon, R. D. Revised Effective Ionic Radii and Systematic Studies of Interatomic Distances in Halides and Chalcogenides. *Acta Cryst A* **1976**, *32*, 751–767.

(50) Nakatsuji, S.; Nambu, Y.; Tonomura, H.; Sakai, O.; Jonas, S.; Broholm, C.; Tsunetsugu, H.; Qiu, Y.; Maeno, Y. Spin Disorder on a Triangular Lattice. *Science* **2005**, *309*, 1697–1700.

(51) Mydosh, J. A. *Spin Glasses: An Experimental Introduction*; CRC Press: London, 1993.

(52) Gallego, S.V.; Tasci, E.S.; Flor, G.; Perez-Mato, J.M.; Aroyo, M.I. Magnetic symmetry in the Bilbao Crystallographic Server: a computer program to provide systematic absences of magnetic neutron diffraction. *J. Appl. Cryst.* **2012**, *45*, 1236-1247.

(53) Chen, J.; Calder, S.; Paddison, J. A. M.; Angelo, G.; Klivansky, L.; Zhang, J.; Cao, H.; Gui, X. $ASb_3Mn_9O_{19}$ (A = K or Rb): New Mn-Based 2D Magnetoplumbites with Geometric and Magnetic Frustration. *Adv. Mater.* **2025**, *37*, 2417906.

(54) Zhao, Z.Y.; Yue, X.Y.; Li, J.Y.; Li, N.; Che, H.L.; Sun, X.F.; He, Z.Z. $Na_5Co_{15.5}Te_6O_{36}$: An S= 1/2 stacked Ising kagome antiferromagnet with a partially disordered ground state. *Phys. Rev. B* **2022**, *105*, 144406.

**Table 1.** Single crystal structure refinement for $(Co_{1-x}Mg_x)_{10}Ge_3O_{16}$ at room temperature.

| Refined x | 0.7 (3)% | 5 (1)% | 9 (1)% | 14 (1)% | 19 (1)% | 29 (2)% |
|---|---|---|---|---|---|---|
| Loading x | 1% | 5% | 10% | 15 | 20% | 30% |
| Temperature (K) | 304 (2) | 303 (2) | 303 (2) | 303 (2) | 303 (2) | 306 (2) |
| F.W. (g/mol) | 1060.76 | 1045.99 | 1031.45 | 1016.10 | 995.91 | 962.33 |
| Space group; Z | $R\bar{3}m$; 3 | $R\bar{3}m$; 3 | $R\bar{3}m$; 3 | $R\bar{3}m$; 3 | $R\bar{3}m$; 3 | $R\bar{3}m$; 3 |
| $a$(Å) | 5.960 (1) | 5.957 (1) | 5.953 (1) | 5.947 (1) | 5.941 (1) | 5.940 (1) |
| $c$(Å) | 28.933 (9) | 28.931 (7) | 28.895 (1) | 28.866 (3) | 28.829 (13) | 28.811 (3) |
| V (Å$^3$) | 889.9 (5) | 889.1 (3) | 886.7 (5) | 884.9 (1) | 881.1 (7) | 880.4 (1) |
| Extinction coefficient | 0.0009(1) | \ | \ |  | \ | \ |
| θ range (°) | 4.011-33.896 | 4.012-35.034 | 4.015-34.976 | 4.019-35.013 | 2.119-35.291 | 4.024-31.083 |
| No. reflections; $R_{int}$ | 4043; 0.1292 | 5710; 0.1470 | 4176; 0.1369 | 4955; 0.0987 | 5966; 0.0891 | 4172; 0.0707 |
| No. independent reflections | 505 | 547 | 544 | 544 | 551 | 404 |
| No. parameters | 30 | 31 | 31 | 31 | 31 | 31 |
| $R_1$: $\omega R_2$ ($I>2\delta(I)$) | 0.0563; 0.0927 | 0.0600; 0.1063 | 0.0647; 0.1121 | 0.0555; 0.0894 | 0.0381; 0.0686 | 0.0334; 0.0570 |
| Goodness of fit | 1.049 | 1.056 | 1.099 | 1.101 | 1.078 | 1.081 |
| Diffraction peak and hole (e$^-$/ Å$^3$) | 2.078; -1.863 | 2.082; -2.622 | 1.887; -2.267 | 1.651; -2.328 | 1.284; -1.536 | 0.955; -1.236 |

**Table 2.** Bond lengths and angles in Co1/Co2/Co3@O$_6$ octahedra of (Co$_{1-x}$Mg$_x$)$_{10}$Ge$_3$O$_{16}$. Note that $d_{ap.}$ is equal to $d_{eq.}$ for Co1.

| | Atom# | 0% | 1% | 5% | 10% | 15% | 20% | 30% |
|---|---|---|---|---|---|---|---|---|
| $d_{ap.}$ (Å) | Co1 | 2.118 (3) | 2.113 (6) | 2.124 (6) | 2.117 (7) | 2.120 (5) | 2.114 (7) | 2.114 (4) |
| | Co2 | 2.116 (3) | 2.125 (6) | 2.126 (6) | 2.123 (7) | 2.122 (4) | 2.127 (9) | 2.122 (4) |
| | Co3 | 2.080 (3) | 2.080 (3) | 2.075 (6) | 2.080 (7) | 2.076 (6) | 2.081 (9) | 2.073 (4) |
| $d_{eq.}$ (Å) | Co1 | 2.118 (3) | 2.113 (6) | 2.124 (6) | 2.117 (7) | 2.120 (5) | 2.114 (7) | 2.114 (4) |
| | Co2 | 2.117 (2) | 2.121 (4) | 2.124 (4) | 2.122 (5) | 2.123 (3) | 2.116 (6) | 2.116 (3) |
| | Co3 | 2.088 (2) | 2.093 (4) | 2.090 (4) | 2.089 (5) | 2.086 (4) | 2.089 (5) | 2.081 (3) |
| $\delta$ (°) | Co1 | 91.0 (1) | 91.1 (2) | 90.9 (3) | 91.0 (3) | 91.2 (2) | 91.0 (3) | 90.9 (1) |
| | Co2 | 89.7 (1) | 89.5 (2) | 89.4 (2) | 89.2 (2) | 89.3 (1) | 89.2 (3) | 89.3 (1) |
| | Co3 | 91.7 (1) | 91.8 (3) | 91.9 (3) | 91.9 (3) | 92.0 (3) | 91.8 (3) | 92.0 (1) |

**Table 3.** Single crystal structure refinement for $(Co_{0.7}Mg_{0.3})_{10}Ge_3O_{16}$ at 107 (2) K.

| | |
|---|---|
| Refined x | 29 (1)% |
| Loading x | 30% |
| Temperature (K) | 107 (2) |
| F.W. (g/mol) | 964.29 |
| Space group; Z | $R$ -3m; 3 |
| $a$(Å) | 5.9330 (2) |
| $c$(Å) | 28.778 (2) |
| V (Å$^3$) | 877.27 (9) |
| θ range (°) | 2.123- 33.721 |
| No. reflections; $R_{int}$ | 5993; 0.0693 |
| No. independent reflections | 490 |
| No. parameters | 31 |
| $R_1$: $\omega R_2$ ($I$>2δ($I$)) | 0.0224; 0.0420 |
| Goodness of fit | 1.085 |
| Diffraction peak and hole (e$^-$/ Å$^3$) | 1.633; -2.260 |

**Figure 1 a.** Crystal structure of $Co_{10}Ge_3O_{16}$. Three different Co sites are marked in dark blue, light blue and cyan, respectively. Ge and O atoms are marked in orange and red. **b.** The Co1/Co2/Co3 sublattices with their separation along the *c* axis and $Co^{2+}$-$Co^{2+}$ bond lengths. **c.** $Co@O_6$ octahedron with oxygen along the apical direction and equatorial direction marked as $O_{ap.}$ and $O_{eq.}$. The Co-O bonds along apical and equatorial directions are represented by $d_{ap.}$ and $d_{eq.}$, while the bond angle of $O_{ap.}$-Co-$O_{eq.}$ is δ. **d.** The trend of $Mg^{2+}$ distribution on Co1, Co2 and Co3 sites with varied x in $(Co_{1-x}Mg_x)_{10}Ge_3O_{16}$.

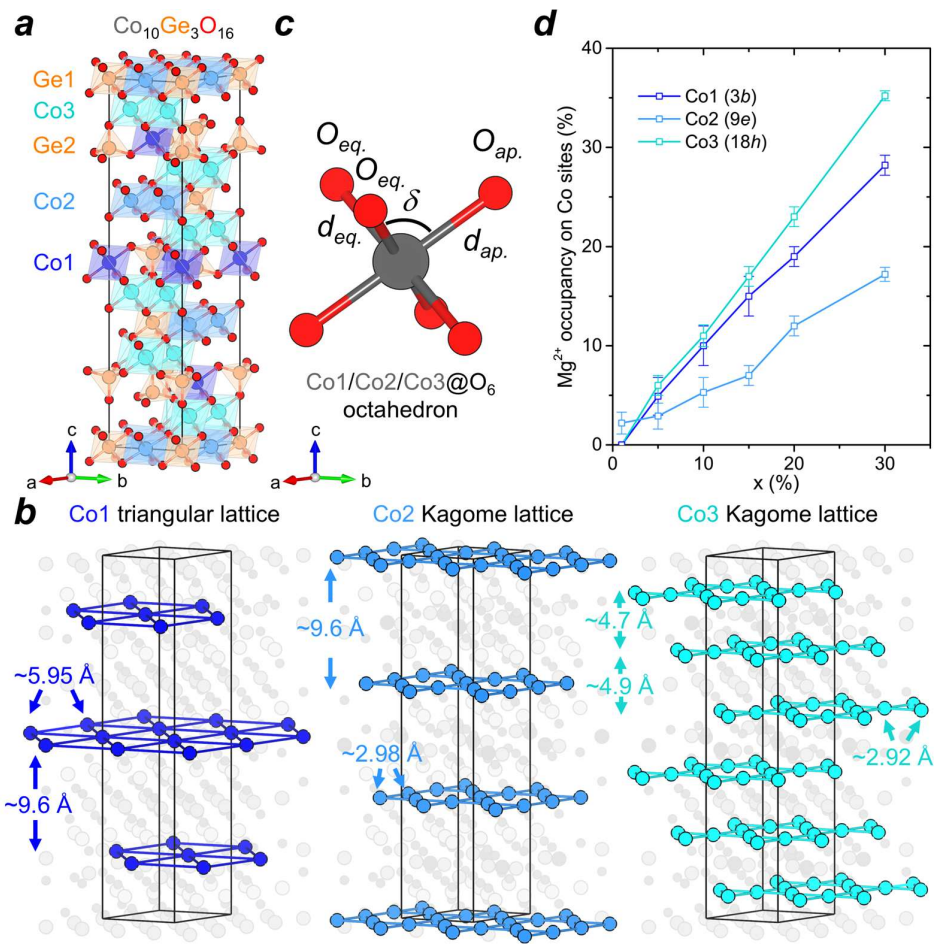

**Figure 2 a. (Main panel)** Powder X-ray diffraction patterns of $(Co_{1-x}Mg_x)_{10}Ge_3O_{16}$ (x = 0%, 1%, 5%, 10%, 15%, 20% and 30%). **(Inset)** Trend of (110) and (0012) peaks. **b.** Changes of lattice parameters and the volume (V) of the unit cell from Le Bail fitting of the powder X-ray diffraction patterns with varied x in $(Co_{1-x}Mg_x)_{10}Ge_3O_{16}$.

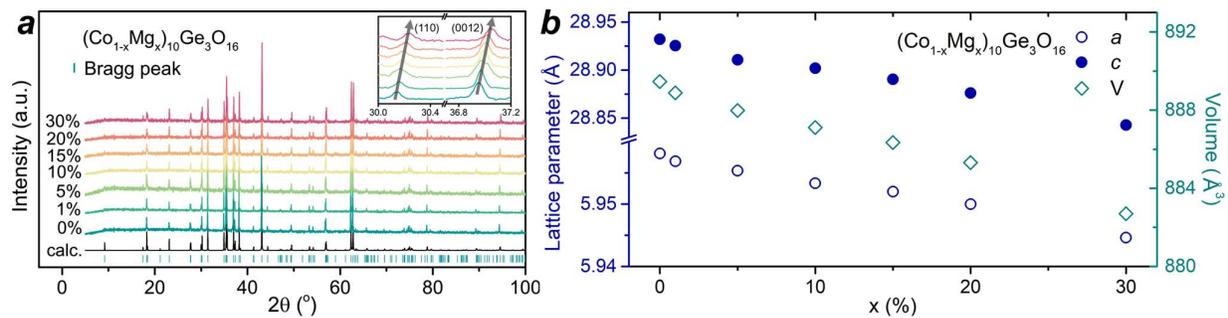

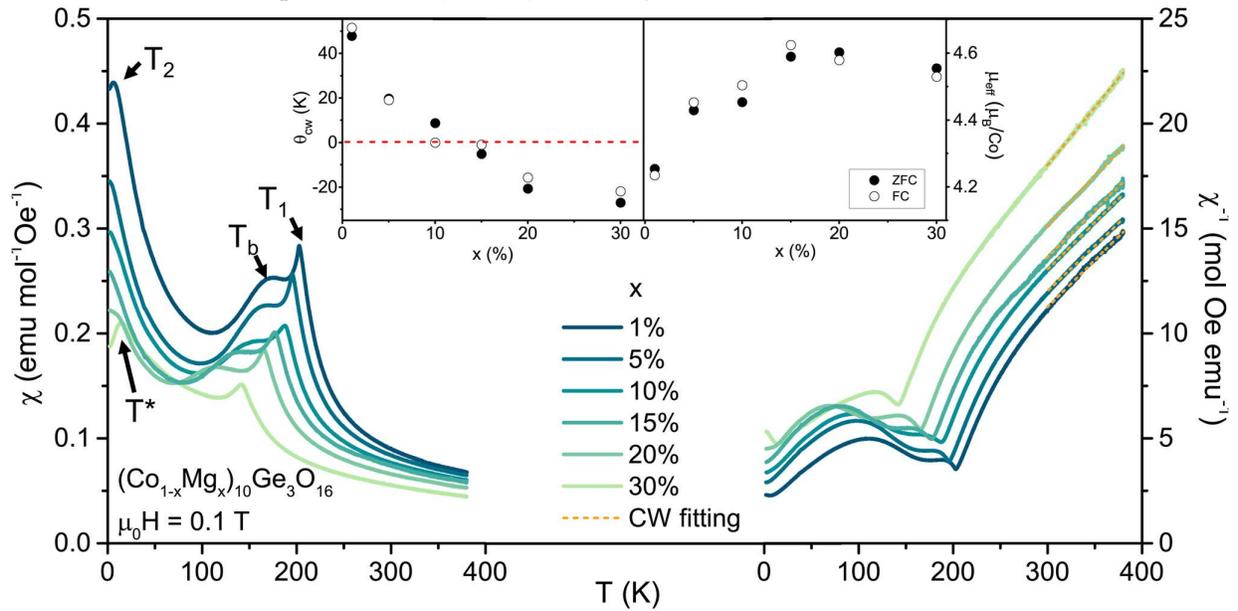

**Figure 3.** Temperature-dependent **(Main panel left)** magnetic susceptibility ($\chi$) and **(Main panel right)** its inverse ($\chi^{-1}$) of $(Co_{1-x}Mg_x)_{10}Ge_3O_{16}$ measured from 2 K to 380 K under an applied magnetic field of 0.1 T. Curie-Weiss fitting is shown in gold. **(Inset)** The fitted Curie-Weiss temperature ($\theta_{CW}$) and effective moment per Co atom ($\mu_{eff}$/Co) varied by x.

**Figure 4 a.** Temperature-dependent magnetic susceptibility ($\chi$) and its inverse ($\chi^{-1}$) of $(Co_{0.7}Mg_{0.3})_{10}Ge_3O_{16}$ measured from 2 K to 380 K under an external magnetic field of 0.1 T. Curie-Weiss fitting is shown in dark yellow line. **b.** AC magnetic susceptibility ($\chi'$) from 2 K to 30 K under AC and DC field of 10 Oe. The AC frequency is varied from 193 Hz to 9984 Hz.

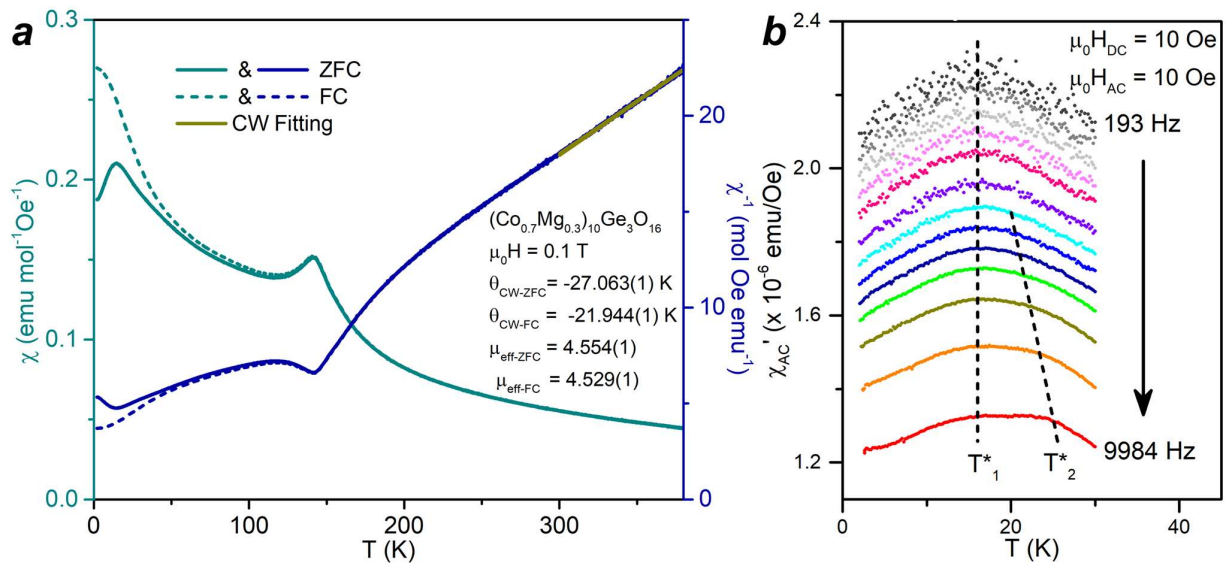

**Figure 5 a. (Main panel)** Heat capacity ($C_p$) of $(Co_{0.7}Mg_{0.3})_{10}Ge_3O_{16}$ measured from 2 K to 200 K under no external magnetic field. The Debye model fitting is shown in dashed line. **(Inset left)** Enlarged view of $C_p$ *vs* T curve from 2 K to 30 K. **(Inset right)** $C_p/T$ *vs* T curve from 2 K to 30 K. **b.** Magnonic contribution to heat capacity ($C_{mag}/T$) and the corresponding entropy change ($\Delta S_{mag}/Co$) from 2 K to 300 K. Dotted lines indicate the temperatures where the lower-temperature entropy change tends to saturate.

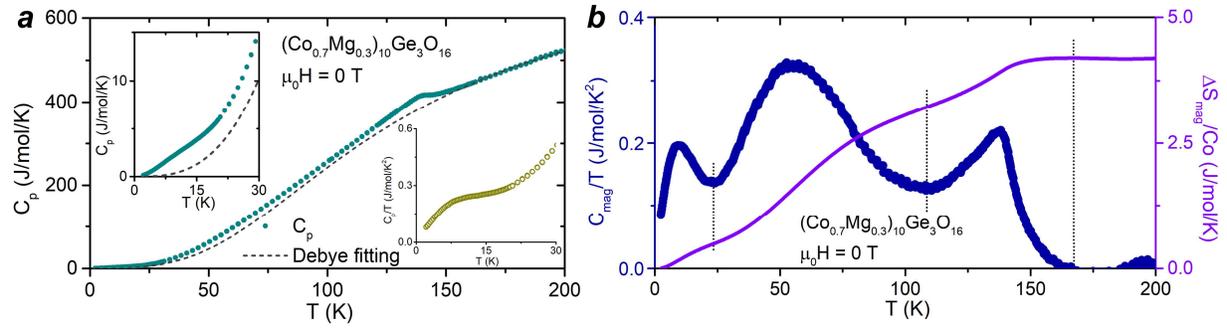

**Figure 6 a.** Neutron powder diffraction patterns of $(Co_{0.7}Mg_{0.3})_{10}Ge_3O_{16}$ at 8 K, 100 K and 300 K. Rietveld fitting, difference and Bragg peak positions are shown in red, blue and green. **b.** The enlarged view of low-Q NPD patterns. Blue asterisks indicate the newly emerged peaks or the peak with enhanced intensity with the *k*-vector of (00 3/2). The space group for all three patterns is *R*-3*m*.

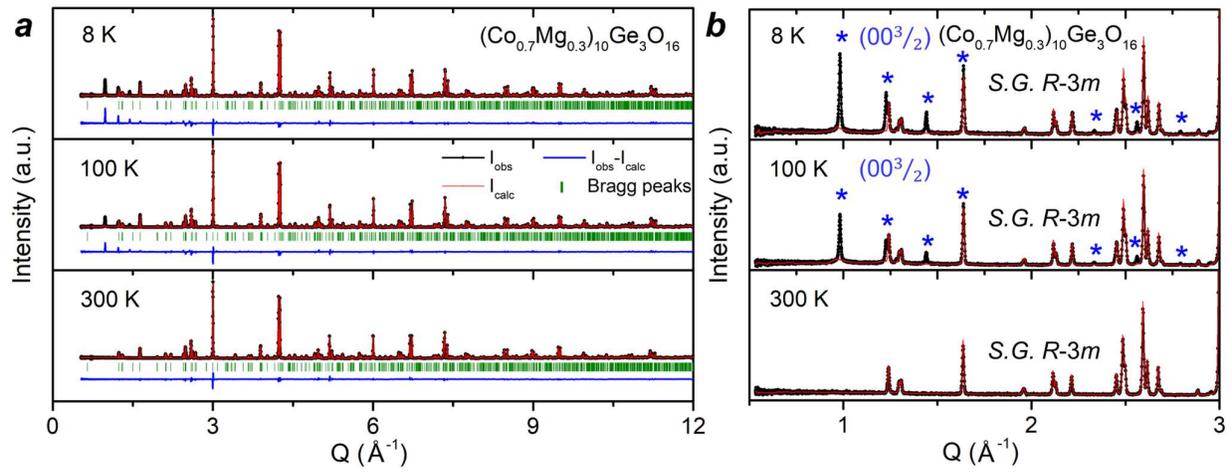

**Figure 7 a.** Hysteresis loops from -9 T to 9 T for $(Co_{1-x}Mg_x)_{10}Ge_3O_{16}$ at 2 K. The virgin curve from 0 to 9 T is marked in red. **b.** The first derivatives of virgin curves. Arrows indicate metamagnetic transitions.

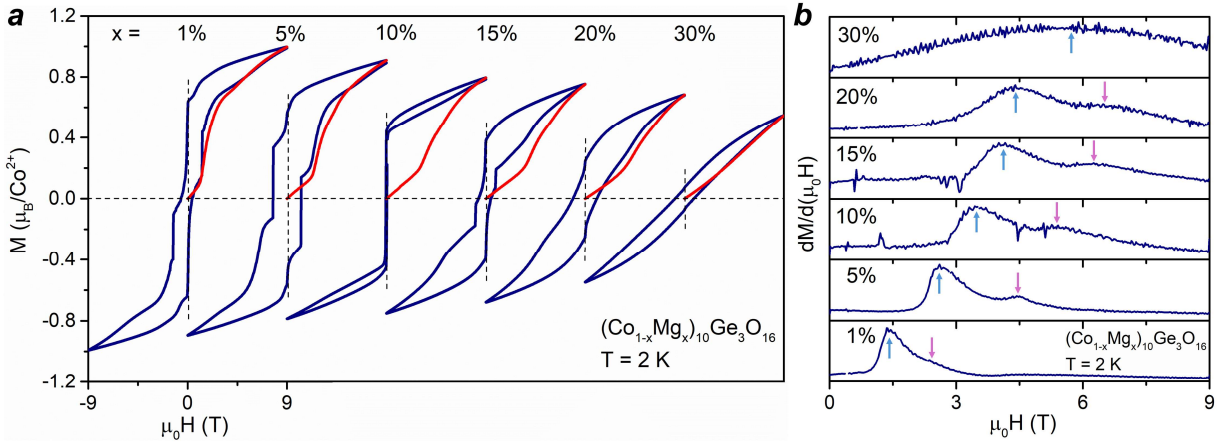

*Supporting Information*

# Site Preference and Possible Coexistence of Antiferromagnetic Order and Magnetic Frustration in $(Co_{1-x}Mg_x)_{10}Ge_3O_{16}$ (0 ≤ x ≤ 30%)

*Gina Angelo,[a] Qiang Zhang,[b] Dylan Correll[a] and Xin Gui[a]\**

[a] Department of Chemistry, University of Pittsburgh, Pittsburgh, PA, 15260, USA
[b] Neutron Scattering Division, Oak Ridge National Laboratory, Oak Ridge, TN, 37831, USA
Address correspondence to E-mail: xig75@pitt.edu

**Table of Contents**

Table S1. Atomic coordinates and equivalent isotropic displacement parameters for $(Co_{1-x}Mg_x)_{10}Ge_3O_{16}$. ($U_{eq}$ is defined as one-third of the trace of the orthogonalized $U_{ij}$ tensor (Å$^2$))

x = 1%

| Atom | Wyck. | Occ. | x | y | z | $U_{eq}$ |
|---|---|---|---|---|---|---|
| Ge1 | 6c | 1 | 2/3 | 1/3 | 0.14408 (5) | 0.005 (1) |
| Ge2 | 3a | 1 | 2/3 | 1/3 | 1/3 | 0.005 (1) |
| Co1 | 3b | 1 | 1/3 | 2/3 | 0.166667 | 0.006 (1) |
| Co2 | 9e | 0.98 (1) | 0.166667 | 1/3 | 1/3 | 0.006 (1) |
| Mg2 | 9e | 0.02 (1) | 0.166667 | 1/3 | 1/3 | 0.006 (1) |
| Co2 | 18h | 1 | 0.5034 (1) | 0.4967 (1) | 0.24776 (4) | 0.006 (1) |
| O1 | 6c | 1 | 2/3 | 1/3 | 0.2058 (3) | 0.005 (2) |
| O2 | 18h | 1 | 0.5020 (5) | 0.004 (1) | 0.1253 (2) | 0.006 (1) |
| O3 | 6c | 1 | 1/3 | 2/3 | 0.2902 (3) | 0.010 (2) |
| O4 | 18h | 1 | 0.8505 (6) | 0.1495 (6) | 0.0379 (2) | 0.010 (1) |

**x = 5%**

| Atom | Wyck. | Occ. | x | y | z | $U_{eq}$ |
|------|-------|------|-----|-----|-----|-----|
| Ge1 | 6c | 1 | 2/3 | 1/3 | 0.14408 (5) | 0.006 (1) |
| Ge2 | 3a | 1 | 2/3 | 1/3 | 1/3 | 0.006 (1) |
| Co1 | 3b | 0.95 (2) | 1/3 | 2/3 | 0.166667 | 0.007 (1) |
| Mg1 | 3b | 0.05 (2) | 1/3 | 2/3 | 0.166667 | 0.007 (1) |
| Co2 | 9e | 0.97 (1) | 0.166667 | 1/3 | 1/3 | 0.007 (1) |
| Mg2 | 9e | 0.03 (1) | 0.166667 | 1/3 | 1/3 | 0.007 (1) |
| Co3 | 18h | 0.94 (1) | 0.5033 (1) | 0.4967 (1) | 0.24779 (4) | 0.006 (1) |
| Mg3 | 18h | 0.06 (1) | 0.5033 (1) | 0.4967 (1) | 0.24779 (4) | 0.006 (1) |
| O1 | 6c | 1 | 2/3 | 1/3 | 0.2060 (3) | 0.008 (2) |
| O2 | 18h | 1 | 0.5028 (5) | 0.006 (1) | 0.1250 (2) | 0.009 (1) |
| O3 | 6c | 1 | 1/3 | 2/3 | 0.2901 (3) | 0.005 (2) |
| O4 | 18h | 1 | 0.8506 (6) | 0.1494 (6) | 0.0381 (2) | 0.009 (1) |

**x = 10%**

| Atom | Wyck. | Occ. | x | y | z | $U_{eq}$ |
|---|---|---|---|---|---|---|
| Ge1 | 6c | 1 | 2/3 | 1/3 | 0.14405 (6) | 0.005 (1) |
| Ge2 | 3a | 1 | 2/3 | 1/3 | 1/3 | 0.006 (1) |
| Co1 | 3b | 0.90 (2) | 1/3 | 2/3 | 0.166667 | 0.008 (1) |
| Mg1 | 3b | 0.10 (2) | 1/3 | 2/3 | 0.166667 | 0.008 (1) |
| Co2 | 9e | 0.95 (2) | 0.166667 | 1/3 | 1/3 | 0.007 (1) |
| Mg2 | 9e | 0.05 (2) | 0.166667 | 1/3 | 1/3 | 0.007 (1) |
| Co3 | 18h | 0.89 (1) | 0.5035 (1) | 0.4965 (1) | 0.24775 (5) | 0.007 (1) |
| Mg3 | 18h | 0.11 (1) | 0.5035 (1) | 0.4965 (1) | 0.24775 (5) | 0.007 (1) |
| O1 | 6c | 1 | 2/3 | 1/3 | 0.2054 (4) | 0.010 (2) |
| O2 | 18h | 1 | 0.5024 (6) | 0.005 (1) | 0.1251 (2) | 0.007 (1) |
| O3 | 6c | 1 | 1/3 | 2/3 | 0.2902 (4) | 0.009 (2) |
| O4 | 18h | 1 | 0.8504 (6) | 0.1496 (6) | 0.0382 (2) | 0.009 (1) |

**x = 15%**

| Atom | Wyck. | Occ. | x | y | z | $U_{eq}$ |
|---|---|---|---|---|---|---|
| Ge1 | 6c | 1 | 2/3 | 1/3 | 0.14411(4) | 0.0040(3) |
| Ge2 | 3a | 1 | 2/3 | 1/3 | 1/3 | 0.0038(4) |
| Co1 | 3b | 0.85(1) | 1/3 | 2/3 | 0.166667 | 0.0056(7) |
| Mg1 | 3b | 0.15(1) | 1/3 | 2/3 | 0.166667 | 0.0056(7) |
| Co2 | 9e | 0.93(1) | 0.166667 | 1/3 | 1/3 | 0.0056(4) |
| Mg2 | 9e | 0.07(1) | 0.166667 | 1/3 | 1/3 | 0.0056(4) |
| Co3 | 18h | 0.83(1) | 0.5034(1) | 0.4966(1) | 0.24775(3) | 0.0058(3) |
| Mg3 | 18h | 0.17(1) | 0.5034(1) | 0.4966(1) | 0.24775(3) | 0.0058(3) |
| O1 | 6c | 1 | 2/3 | 1/3 | 0.2060(3) | 0.008(2) |
| O2 | 18h | 1 | 0.5020(5) | 0.004(1) | 0.1252(2) | 0.006(1) |
| O3 | 6c | 1 | 1/3 | 2/3 | 0.2901(2) | 0.005(2) |
| O4 | 18h | 1 | 0.8504(5) | 0.1496(5) | 0.0382(2) | 0.008(1) |

**x = 20%**

| Atom | Wyck. | Occ. | x | y | z | $U_{eq}$ |
|------|-------|------|-----|-----|-----|------|
| Ge1 | 6c | 1 | 2/3 | 1/3 | 0.14413 (3) | 0.005 (1) |
| Ge2 | 3a | 1 | 2/3 | 1/3 | 1/3 | 0.005 (1) |
| Co1 | 3b | 0.81 (1) | 1/3 | 2/3 | 0.166667 | 0.007 (1) |
| Mg1 | 3b | 0.19 (1) | 1/3 | 2/3 | 0.166667 | 0.007 (1) |
| Co2 | 9e | 0.88 (1) | 0.166667 | 1/3 | 1/3 | 0.007 (1) |
| Mg2 | 9e | 0.12 (1) | 0.166667 | 1/3 | 1/3 | 0.007 (1) |
| Co3 | 18h | 0.77 (1) | 0.5033 (1) | 0.4967 (1) | 0.2478 (1) | 0.006 (1) |
| Mg3 | 18h | 0.23 (1) | 0.5033 (1) | 0.4967 (1) | 0.2478 (1) | 0.006 (1) |
| O1 | 6c | 1 | 2/3 | 1/3 | 0.2055 (1) | 0.006 (1) |
| O2 | 18h | 1 | 0.5023 (3) | 0.0046 (6) | 0.1249 (1) | 0.007 (1) |
| O3 | 6c | 1 | 1/3 | 2/3 | 0.2903 (1) | 0.007 (1) |
| O4 | 18h | 1 | 0.8505 (3) | 0.1495 (3) | 0.0382 (1) | 0.007 (1) |

**x = 30% at 306 K**

| Atom | Wyck. | Occ. | x | y | z | $U_{eq}$ |
|---|---|---|---|---|---|---|
| Ge1 | 6c | 1 | 2/3 | 1/3 | 0.14412 (3) | 0.0043 (2) |
| Ge2 | 3a | 1 | 2/3 | 1/3 | 1/3 | 0.0043 (3) |
| Co1 | 3b | 0.72(1) | 1/3 | 2/3 | 0.166667 | 0.0056 (6) |
| Mg1 | 3b | 0.28 (1) | 1/3 | 2/3 | 0.166667 | 0.0056 (6) |
| Co2 | 9e | 0.828 (7) | 0.166667 | 1/3 | 1/3 | 0.0063 (3) |
| Mg2 | 9e | 0.172 (7) | 0.166667 | 1/3 | 1/3 | 0.0063 (3) |
| Co3 | 18h | 0.648 (5) | 0.50314 (8) | 0.49686 (8) | 0.24783 (3) | 0.0059 (3) |
| Mg3 | 18h | 0.352 (5) | 0.50314 (8) | 0.49686 (8) | 0.24783 (3) | 0.0059 (3) |
| O1 | 6c | 1 | 2/3 | 1/3 | 0.2058 (2) | 0.005 (1) |
| O2 | 18h | 1 | 0.5024 (3) | 0.0047 (6) | 0.1250 (1) | 0.0055 (7) |
| O3 | 6c | 1 | 1/3 | 2/3 | 0.2900 (2) | 0.007 (1) |
| O4 | 18h | 1 | 0.8505 (3) | 0.1495 (3) | 0.0380 (1) | 0.008 (2) |

**x = 30% at 107 K**

| Atom | Wyck. | Occ. | x | y | z | $U_{eq}$ |
|------|-------|------|---|---|---|----------|
| Ge1 | 6c | 1 | 2/3 | 1/3 | 0.14412 (2) | 0.0035 (1) |
| Ge2 | 3a | 1 | 2/3 | 1/3 | 1/3 | 0.0030 (2) |
| Co1 | 3b | 0.74(1) | 1/3 | 2/3 | 0.166667 | 0.0043 (3) |
| Mg1 | 3b | 0.26 (1) | 1/3 | 2/3 | 0.166667 | 0.0043 (3) |
| Co2 | 9e | 0.819 (5) | 0.166667 | 1/3 | 1/3 | 0.0034 (1) |
| Mg2 | 9e | 0.181 (5) | 0.166667 | 1/3 | 1/3 | 0.0034 (1) |
| Co3 | 18h | 0.658 (4) | 0.50322 (4) | 0.49678 (4) | 0.24789 (2) | 0.0041 (2) |
| Mg3 | 18h | 0.342 (4) | 0.50322 (4) | 0.49678 (4) | 0.24789 (2) | 0.0041 (2) |
| O1 | 6c | 1 | 2/3 | 1/3 | 0.2060 (1) | 0.005 (1) |
| O2 | 18h | 1 | 0.5024 (2) | 0.0049 (4) | 0.1250 (1) | 0.0050 (4) |
| O3 | 6c | 1 | 1/3 | 2/3 | 0.2900 (1) | 0.005 (1) |
| O4 | 18h | 1 | 0.8499 (1) | 0.1501 (2) | 0.0383 (1) | 0.005 (1) |

**Table S2.** Anisotropic thermal displacement parameters for $(Co_{1-x}Mg_x)_{10}Ge_3O_{16}$.

$x = 1\%$

| Atom | U11 | U22 | U33 | U23 | U13 | U12 |
|------|-----|-----|-----|-----|-----|-----|
| Ge1 | 0.0038(4) | 0.0038(4) | 0.0071(7) | 0 | 0 | 0.0019(2) |
| Ge2 | 0.0043(6) | 0.0043(6) | 0.007(1) | 0 | 0 | 0.0021(3) |
| Co1 | 0.0066(8) | 0.0066(8) | 0.005(1) | 0 | 0 | 0.0033(4) |
| Co2 | 0.0035(6) | 0.0058(8) | 0.0085(8) | 0.0014(6) | 0.0007(3) | 0.0029(4) |
| Mg2 | 0.0035(6) | 0.0058(8) | 0.0085(8) | 0.0014(6) | 0.0007(3) | 0.0029(4) |
| Co3 | 0.0063(4) | 0.0063(4) | 0.0072(5) | 0.0004(2) | -0.0004(2) | 0.0034(5) |

$x = 5\%$

| Atom | U11 | U22 | U33 | U23 | U13 | U12 |
|------|-----|-----|-----|-----|-----|-----|
| Ge1 | 0.0060(5) | 0.0060(5) | 0.0049(6) | 0 | 0 | 0.0030(2) |
| Ge2 | 0.0057(6) | 0.0057(6) | 0.0058(8) | 0 | 0 | 0.0028(3) |
| Co1 | 0.008(1) | 0.008(1) | 0.007(1) | 0 | 0 | 0.0039(5) |
| Mg1 | 0.008(1) | 0.008(1) | 0.007(1) | 0 | 0 | 0.0039(5) |
| Co2 | 0.0055(6) | 0.0072(8) | 0.0088(7) | 0.0010(6) | 0.0005(3) | 0.0036(4) |
| Mg2 | 0.0055(6) | 0.0072(8) | 0.0088(7) | 0.0010(6) | 0.0005(3) | 0.0036(4) |
| Co3 | 0.0066(5) | 0.0066(5) | 0.0066(5) | 0.0003(2) | -0.0003(2) | 0.0040(5) |
| Mg3 | 0.0066(5) | 0.0066(5) | 0.0066(5) | 0.0003(2) | -0.0003(2) | 0.0040(5) |

$x = 10\%$

| Atom | U11 | U22 | U33 | U23 | U13 | U12 |
|------|-----|-----|-----|-----|-----|-----|
| Ge1 | 0.0052(5) | 0.0052(5) | 0.0057(7) | 0 | 0 | 0.0026(3) |
| Ge2 | 0.0059(7) | 0.0059(7) | 0.005(1) | 0 | 0 | 0.0030(4) |
| Co1 | 0.010(1) | 0.010(1) | 0.004(1) | 0 | 0 | 0.0050(6) |
| Mg1 | 0.010(1) | 0.010(1) | 0.004(1) | 0 | 0 | 0.0050(6) |
| Co2 | 0.0064(7) | 0.0072(9) | 0.0089(9) | 0.0006(7) | 0.0003(3) | 0.0036(5) |
| Mg2 | 0.0064(7) | 0.0072(9) | 0.0089(9) | 0.0006(7) | 0.0003(3) | 0.0036(5) |
| Co3 | 0.0080(5) | 0.0080(5) | 0.0064(6) | 0.0002(2) | -0.0002(2) | 0.0040(6) |
| Mg3 | 0.0080(5) | 0.0080(5) | 0.0064(6) | 0.0002(2) | -0.0002(2) | 0.0040(6) |

**x = 15%**

| Atom | U11 | U22 | U33 | U23 | U13 | U12 |
|------|-----|-----|-----|-----|-----|-----|
| Ge1 | 0.0042(4) | 0.0042(4) | 0.0034(5) | 0 | 0 | 0.0021(2) |
| Ge2 | 0.0037(5) | 0.0037(5) | 0.004(4) | 0 | 0 | 0.0018(3) |
| Co1 | 0.0056(9) | 0.0056(9) | 0.006(1) | 0 | 0 | 0.0028(4) |
| Mg1 | 0.0056(9) | 0.0056(9) | 0.006(1) | 0 | 0 | 0.0028(4) |
| Co2 | 0.0049(6) | 0.0055(7) | 0.0067(6) | 0.0006(5) | 0.0003(2) | 0.0027(3) |
| Mg2 | 0.0049(6) | 0.0055(7) | 0.0067(6) | 0.0006(5) | 0.0003(2) | 0.0027(3) |
| Co3 | 0.0068(4) | 0.0068(4) | 0.0050(5) | 0.0001(2) | -0.0001(2) | 0.0043(5) |
| Mg3 | 0.0068(4) | 0.0068(4) | 0.0050(5) | 0.0001(2) | -0.0001(2) | 0.0043(5) |

**x = 20%**

| Atom | U11 | U22 | U33 | U23 | U13 | U12 |
|------|-----|-----|-----|-----|-----|-----|
| Ge1 | 0.0053(3) | 0.0053(3) | 0.0038(4) | 0 | 0 | 0.0027(1) |
| Ge2 | 0.0048(4) | 0.0048(4) | 0.0047(5) | 0 | 0 | 0.0024(2) |
| Co1 | 0.0077(6) | 0.077(6) | 0.0055(9) | 0 | 0 | 0.0038(3) |
| Mg1 | 0.077(6) | 0.0077(6) | 0.0055(9) | 0 | 0 | 0.0038(3) |
| Co2 | 0.0059(4) | 0.0075(5) | 0.0066(5) | 0.0003(4) | 0.0002(1) | 0.0037(3) |
| Mg2 | 0.0059(4) | 0.0075(5) | 0.0066(5) | 0.0003(4) | 0.0002(1) | 0.0037(3) |
| Co3 | 0.0066(3) | 0.0066(3) | 0.0061(4) | -0.0002(1) | 0.0002(1) | 0.0037(3) |
| Mg3 | 0.0066(3) | 0.0066(3) | 0.0061(4) | -0.0002(1) | 0.0002(1) | 0.0037(3) |

**x = 30% at 306K**

| Atom | U11 | U22 | U33 | U23 | U13 | U12 |
|------|-----|-----|-----|-----|-----|-----|
| Ge1 | 0.0046(3) | 0.0046(3) | 0.0039(4) | 0 | 0 | 0.0023(2) |
| Ge2 | 0.0045(4) | 0.0045(4) | 0.0040(5) | 0 | 0 | 0.0022(2) |

| Atom | U11 | U22 | U33 | U23 | U13 | U12 |
|------|-----|-----|-----|-----|-----|-----|
| Co1 | 0.0063(7) | 0.0063(7) | 0.0042(9) | 0 | 0 | 0.0031(4) |
| Mg1 | 0.0063(7) | 0.0063(7) | 0.0042(9) | 0 | 0 | 0.0031(4) |
| Co2 | 0.0058(5) | 0.0063(6) | 0.0069(5) | 0.0011(4) | 0.0005(2) | 0.0032(3) |
| Mg2 | 0.0058(5) | 0.0063(6) | 0.0069(5) | 0.0011(4) | 0.0005(2) | 0.0032(3) |
| Co3 | 0.0063(4) | 0.0063(4) | 0.0062(4) | 0.0002(2) | -0.0002(2) | 0.0041(4) |
| Mg3 | 0.0063(4) | 0.0063(4) | 0.0062(4) | 0.0002(2) | -0.0002(2) | 0.0041(4) |

**x = 30% at 107K**

| Atom | U11 | U22 | U33 | U23 | U13 | U12 |
|------|-----|-----|-----|-----|-----|-----|
| Ge1 | 0.0033(2) | 0.0033(2) | 0.0037(2) | 0 | 0 | 0.0017(1) |
| Ge2 | 0.0031(2) | 0.0031(2) | 0.0030(3) | 0 | 0 | 0.0015(1) |
| Co1 | 0.0039(4) | 0.0039(4) | 0.0049(5) | 0 | 0 | 0.0020(2) |
| Mg1 | 0.0039(4) | 0.0039(4) | 0.0049(5) | 0 | 0 | 0.0020(2) |
| Co2 | 0.0029(2) | 0.0035(3) | 0.0040(3) | 0.0002(2) | 0.0001(1) | 0.0017(1) |
| Mg2 | 0.0029(2) | 0.0035(3) | 0.0040(3) | 0.0002(2) | 0.0001(1) | 0.0013(1) |
| Co3 | 0.0044(2) | 0.0044(2) | 0.0038(2) | -0.0001(1) | 0.0001(1) | 0.0025(1) |
| Mg3 | 0.0044(2) | 0.0044(2) | 0.0038(2) | -0.0001(1) | 0.0001(1) | 0.0025(1) |

**Figure S1.** Magnetic susceptibility of $(Co_{1-x}Mg_x)_{10}Ge_3O_{16}$ (x = 1%, 5%, 10%, 15%, 20% and 30%) (red, orange, green, blue, purple, pink, respectively) under 0.1T and field-cooling (FC) protocol on the left-hand side of figure with inverse magnetic susceptibility on the right. The inset on the left shows the trend of decreasing CW temperature with increasing Mg% for both ZFC (filled black circles) and FC (empty black circles). The dashed red line indicates 0 K, separating dominant antiferromagnetic interactions from dominant ferromagnetic interactions. The right inset depicts the trend between Mg% and effective moment per Co atom for both ZFC (filled black circles) and FC (empty black circles).

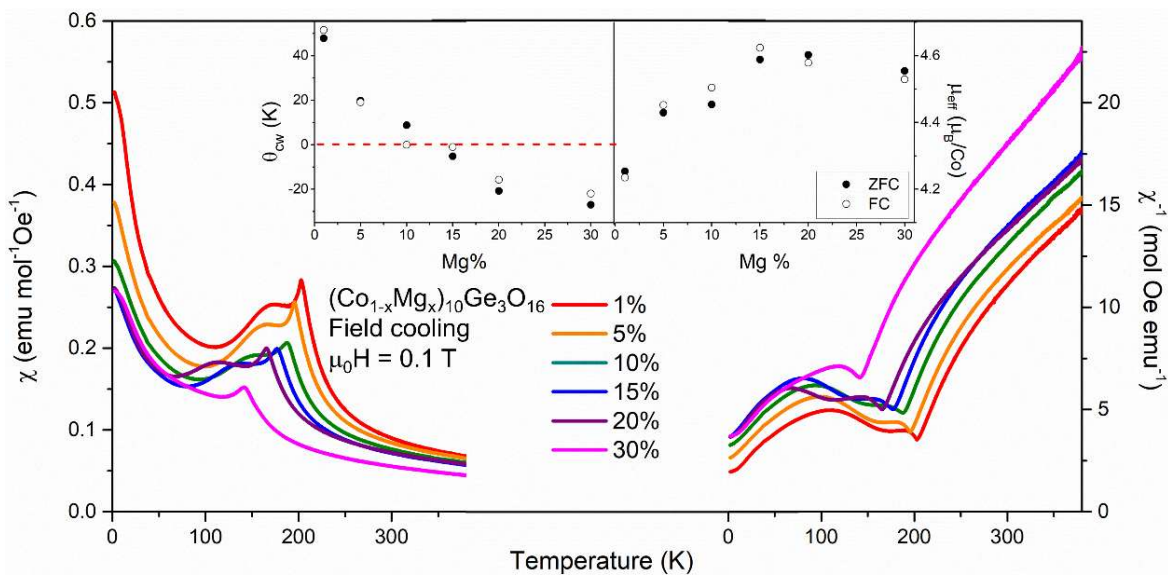

**Figure S2.** The first derivative of AC magnetic susceptibility of $(Co_{0.7}Mg_{0.3})_{10}Ge_3O_{16}$. A red line and black arrow were added to clarify that there is a kink that emerges from 1389 Hz. It is most easily noticed at the highest frequency of 9984 Hz, but is less noticeable in the lower frequencies.

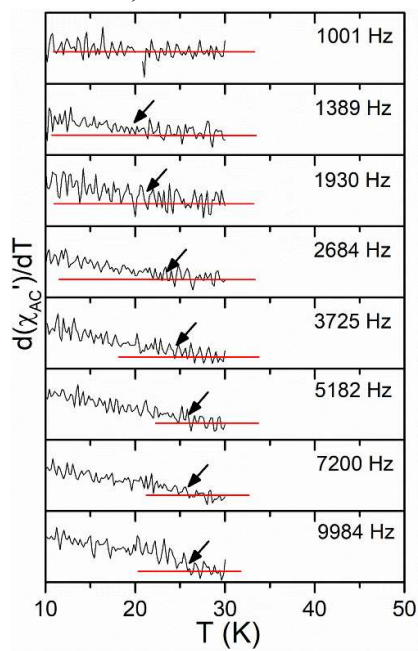

**Figure S3.** Possible magnetic space group of $(Co_{0.7}Mg_{0.3})_{10}Ge_3O_{16}$ obtained from Bilbao server [Gallego, S. V., Perez-Mato, J. M., Elcoro, L., Tasci, E., de la Flor, G., & Aroyo, M. I. (2012). Magnetic symmetry in the Bilbao Crystallographic Server: a computer program to provide systematic absences of magnetic neutron diffraction. *Journal of Applied Crystallography*, *45*(6), 1236–1247.]

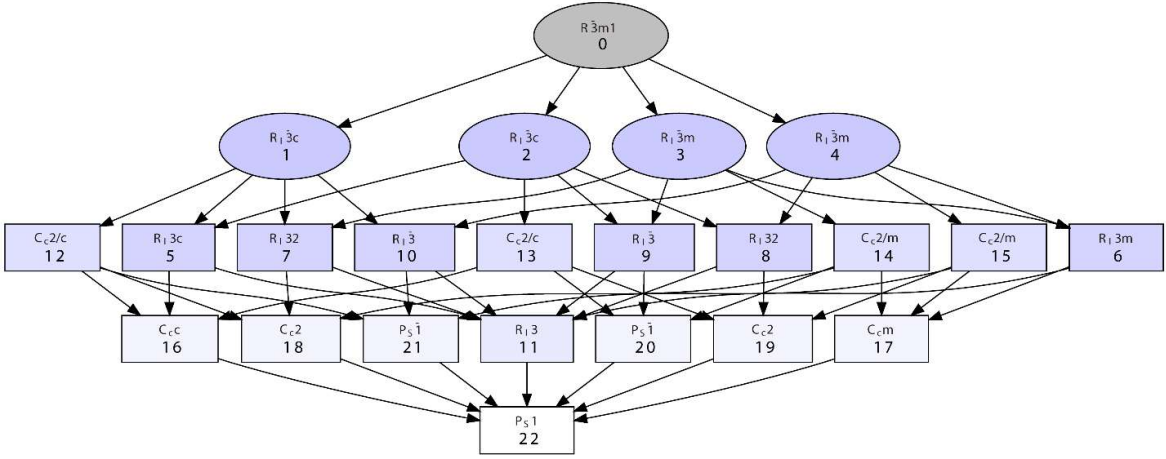

**Figure S4.** Proposed evolution of magnetic order/interaction in $Co_{10}Ge_3O_{16}$ and $(Co_{1-x}Mg_x)_{10}Ge_3O_{16}$ where three different Co sites are marked in dark blue, light blue and cyan, respectively. Ge and O atoms are marked in orange and red. NM, AFM, SRC, SRO and SM stand for non-magnetic, antiferromagnetic, short-range correlation, short-range order and static moment. Potential structural phase transition from *R-3m* to *C2/m* exists below $T_1$. However, since we did not observe the *C2/m* structure below $T_1$ for x = 30%, it is still unsure which low-temperature crystal structure $1\% \leq x < 30\%$

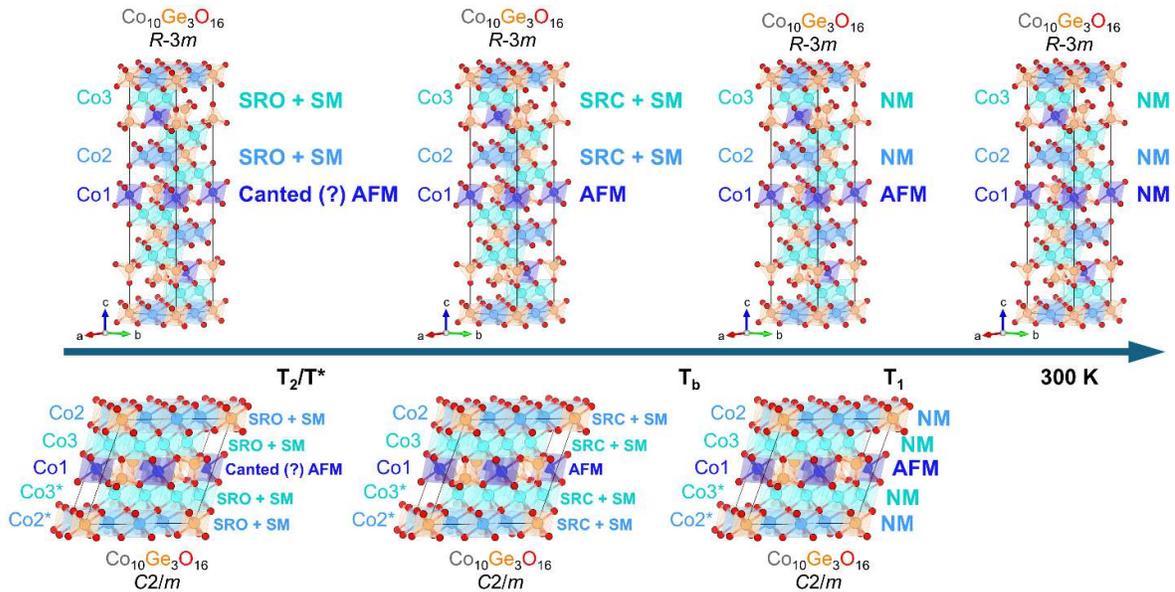

will adopt.

**Figure S5**. Hysteresis of $(Co_{1-x}Mg_x)_{10}Ge_3O_{16}$ (x = 1%, 5%, 10%, 15%, 20%, 30%) at **a.** 40 K, **b.** 150 K, and **c.** 220 K. **d.** Virgin magnetization curve from 0-9 T at 300 K for x= 1%, 5%, 10%, 15%, 20%, 30%.

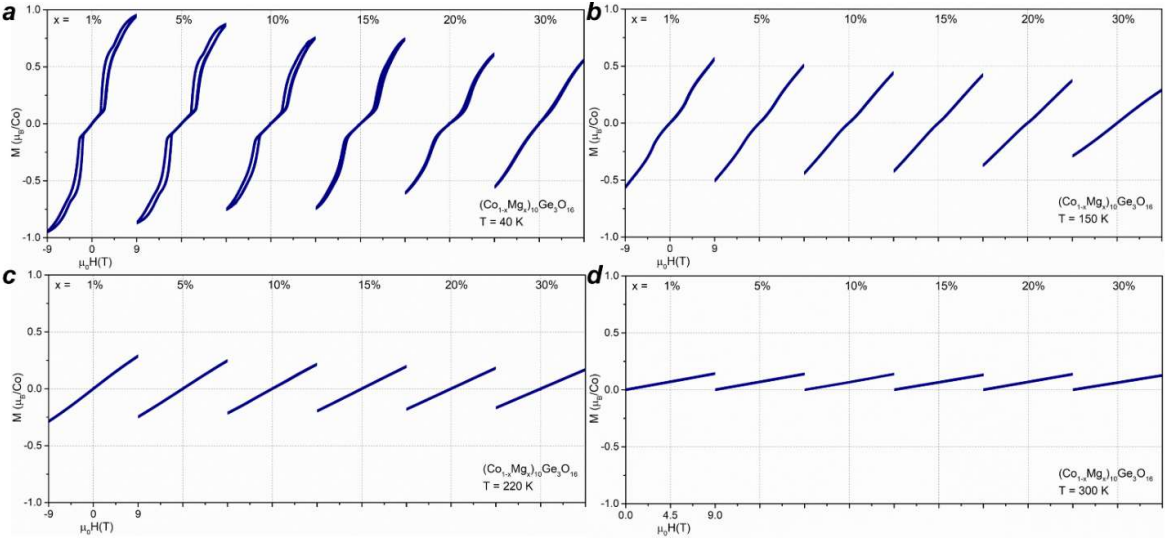

**Figure S6**. Comparison of the hysteresis loops and virgin curves of $(Co_{1-x}Mg_x)_{10}Ge_3O_{16}$ (x = 1%, 5%, 10%, 15%, 20%, 30%) with a batch size of 200 mg in red dashed line and 2 grams in blue solid line.

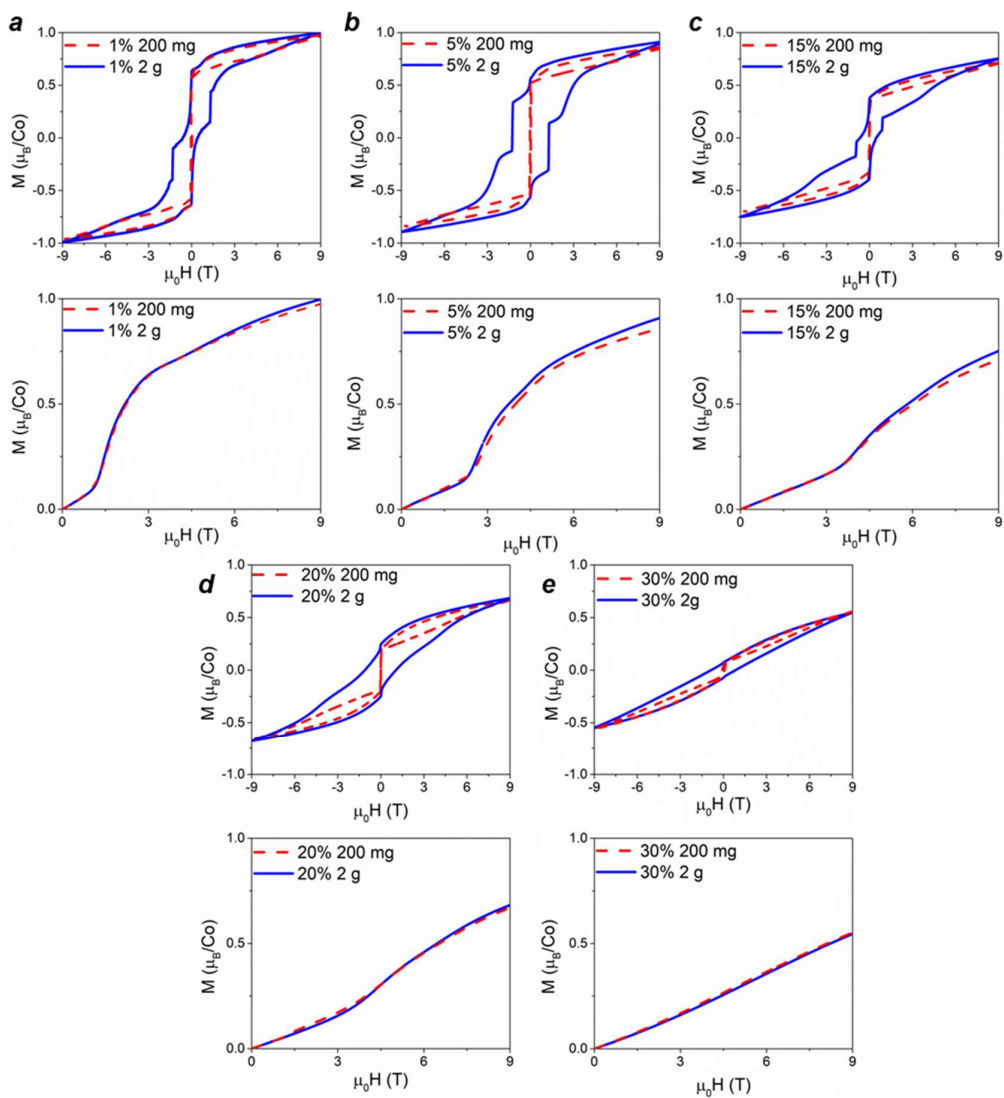

The top of each panel shows the hysteresis loops while the bottom are virgin curves.